\documentclass[reqno]{amsart}

\usepackage{amsmath, amssymb, nccmath}
\usepackage{amsthm} 
\usepackage{enumerate}
\usepackage{latexsym,amscd,amsfonts}
\usepackage{graphicx,float}
\usepackage{color}

\usepackage{graphics}
\usepackage{tikz,tkz-tab}

\def \R {\mathbb R}

\def \N {\mathbb N}

\newcommand{\U}{\mathsf{U}}
\newcommand{\V}{\mathsf{V}}

\newcommand{\E}{\mathbf{E}}
\newcommand{\HH}{\mathbf{H}}

\newcommand{\ee}{\mathrm{e}}
\newcommand{\ii}{\mathrm{i}}
\newcommand{\dd}{\mathrm{d}}

\newcommand{\cc}{\mathsf{c}}

\newtheorem{lemma}{Lemma}
\newtheorem{defi}{Definition}
\newtheorem{theoreme}{Theorem}
\newtheorem{prop}{Proposition}

\newenvironment{demo}{\noindent \textit{Proof:}}{\begin{flushright} $\Box$ \end{flushright}}

\title[Multiple wells Airy-Schr\"odinger operator]{Multiple wells Airy-Schr\"odinger operator and integrated density of states of the periodic Airy-Schr\"odinger operator}
\author{H. Boumaza}
\address{H. Boumaza, Universit\'e Sorbonne Paris Nord, LAGA, CNRS, UMR 7539,  F-93430, Villetaneuse, France}
\author{O. Lafitte}
\address{O. Lafitte, Université Sorbonne Paris Nord, LAGA, CNRS, UMR 7539,  F-93430, Villetaneuse, France} 

\date{}

\begin{document}

\begin{abstract}
We compute an explicit formula for the integrated density of states of the periodic Airy-Schr\"odinger operator on the real line. For this purpose, we study precisely the spectrum of the restriction of the periodic Airy-Schr\"odinger operator to a finite number of periods. We prove that all the eigenvalues of this restriction are in the spectral bands of the periodic Airy-Schr\"odinger operator and none of them are in its spectral gaps. We count exactly the number of eigenvalues in each of the spectral bands, this number being equal to the number of periods to which the periodic operator is restricted, plus one. Note that our results depends on a semiclassical parameter but are not stated in the semiclassical limit. They are valid for values of the semiclassical larger than explicit constants. 
\end{abstract}

\maketitle


\section{The model and main results}

In this article, we aim at studying the integrated density of states of the periodic Airy-Schr\"odinger operator studied in \cite{BL}. Along with the explicit computation of this integrated density of states, we determine the localization of the eigenvalues of the restricted operator which is now defined.

\subsection{The multiple wells Airy-Schr\"odinger operator}

 Let $N\geq 0$ be an integer and let $2L_0\in \R_+^*$ be a characteristic length modelling the distance between two ions in a one dimensional finite lattice of $2N+1$ ions. The motion of electrons in this finite lattice can be studied through the following Schr\"odinger operator acting on the subspace of the Sobolev space $\mathsf{H}^2(\mathbb{R})$, 
\[
D(H_{2N+1})=\{\psi\in \mathsf{H}^2(\mathbb{R})\ |\ \psi(-(2N+1)L_0)=\psi((2N+1)L_0) \}
\]
by
\begin{equation}\label{eq_def_HM}
H_{2N+1}= -\frac{\hbar^2}{2m} \frac{\mathrm{d}^2 }{\mathrm{d}x^2} + V_{2N+1}, 
\end{equation}
where $\hbar$ is the reduced Planck constant and $V_{2N+1}$ is the multiplication operator by : 
\begin{equation}\label{eq_def_V2N1}
V_{2N+1} : x\mapsto \sum_{k=-N}^N V(x-2kL_0) 
\end{equation}
where $V\ :\ \R \to \R$ is defined by
\begin{equation}\label{eq_def_V}
V(x)=\left\lbrace \begin{array}{ccl}
                  V_0\left( \mfrac{|x|}{L_0} -1\right) & \mbox{ if } & x \in [-L_0,L_0] \\
		  0 & \mbox{ elsewhere} &
                  \end{array}
 \right. ,
\end{equation}
 $V_0\in \R_+^*$ being a reference potential. The potential $V_{2N+1}$ is continuous on $\R$ and continuously differentiable everywhere except at its minima and maxima points. Note that $V_{2N+1}$ is an even function.
\bigskip

\noindent We recall the definition of the periodic Airy-Schr\"odinger operator as defined in \cite{BL}. Let
\begin{equation}\label{def_H_periodic}
H= -\frac{\hbar^2}{2m} \frac{\mathrm{d}^2 }{\mathrm{d}x^2} + \tilde{V}, 
\end{equation}
where $\tilde{V}$ is the $2L_0$-periodic function equal to $V$ on $[-L_0,L_0]$.  The operator $H$ has purely absolutely continuous spectrum and this spectrum is a union of closed intervals (see \cite[XIII]{RS4}):  
\begin{equation}\label{def_spectral_bands}
\sigma(H)= \bigcup_{p\geq 0} \left[ E_{\mathrm{min}}^p, E_{\mathrm{max}}^p \right].
\end{equation}
For $p\geq 0$,  $[ E_{\mathrm{min}}^p, E_{\mathrm{max}}^p ]$ is called the $p$-th spectral band, $(E_{\mathrm{max}}^p,E_{\mathrm{min}}^{p+1} )$ the $p$-th spectral gap and $E_{\mathrm{min}}^p$ and $E_{\mathrm{max}}^p$ are called the spectral band edges. The band edges are characterized through the theory of Bloch decomposition for periodic Schr\"odinger operators.  More precisely, let $\omega \in [-L_0,L_0]$. Consider the restriction $H(\omega)$ of $H$ to $\mathsf{H}^2([-L_0,L_0])$, the Sobolev space of functions $\psi\in \mathsf{H}^2(\R)$ which satisfy 
\begin{equation}
\label{eq_boundary_cond} \psi(L_0)=\ee^{\ii (\frac{\pi}{L_0}\omega + \pi)} \psi(-L_0)\quad \mbox{and} \quad \psi'(L_0)=\ee^{\ii (\frac{\pi}{L_0}\omega + \pi)} \psi'(-L_0).
 \end{equation}
The operator $H(\omega)$ is self-adjoint, Hilbert-Schmidt and thus compact. Its spectrum consists only on isolated eigenvalues and we set
$$\sigma(H(-L_0)):=  \{ E_{\mathrm{min}}^0, E_{\mathrm{max}}^1, E_{\mathrm{min}}^2, E_{\mathrm{max}}^3,\ldots \} \mbox{ and } \sigma(H(0)):= \{ E_{\mathrm{max}}^0, E_{\mathrm{min}}^1, E_{\mathrm{max}}^2, E_{\mathrm{min}}^3,\ldots \}.$$

\noindent In order to describe the spectrum of the operator $H_{2N+1}$, one considers the equation:
\begin{equation}\label{eq_schrodinger}
 H_{2N+1} \psi=E\psi,\quad E\in \R,\ \psi\in D(H_{2N+1}).
\end{equation}
After rescaling and translating, this equation is equivalent to an Airy equation on each interval $]k L_0, (k+1)L_0[$, for $k\in \{ -(2N+1),\ldots, 2N \}$. Thus we introduce $u$ and $v$, the canonical solutions of the Airy equation, satisfying
$$u(0)=1,\ u'(0)=0\quad \mbox{ and }\quad v(0)=0,\ v'(0)=1.$$
In particular, the Wronskian of $u$ and $v$, $uv'-u'v$, is constant and equal to $1$. Both $u$ and $v$ are analytic functions on $\R$. Moreover, $u$ is strictly decreasing and negative on $[0,+\infty[$ and $v$ is strictly increasing and positive on $(0,+\infty)$. Thus, the zeroes of $u$, $v$ and their derivatives are all non-positive real numbers. 
\vskip 2mm

\noindent \textbf{Notation.} We denote by $\{ -c_{2j+1} \}_{j \geq 0} \cup \{ 0\} $ and  $\{ -c_{2j} \}_{j \geq 0}$ the set of the zeroes of respectively $v$ and $v'$ arranged in decreasing order.
\vskip 2mm

\noindent \textbf{Notation.} Let
\begin{equation}\label{def_param_c}
\cc =\left(\frac{2mL_0^2 V_0}{\hbar^2}\right)^{\frac13}.            
\end{equation}
As we will see in (\ref{eq_schrodinger_rescaled}), this parameter corresponds to the heights of the potential barrier. It will play the role of a semiclassical parameter and the semiclassical limit corresponds to $\cc$ tends to $+\infty$.

\vskip 3mm

\noindent We now state the first result of this article which describe the spectrum of $H_{2N+1}$ and compare it to the spectrum of the periodic Airy-Schr\"odinger operator in the range of the potential, the interval $[-V_0,0]$.

\begin{theoreme}\label{thm_spectre}
\begin{enumerate}
 \item The spectrum of $H_{2N+1}$ is equal to its point spectrum and
\begin{equation}\label{eq_thm_spectre_gaps}
\sigma(H_{2N+1}) \cap [-V_0,0] \subset \sigma(H)\cap [-V_0,0].
\end{equation}
\item For each $p$ in $\N$, for every $\cc \geq c_{p}$ and every $i\in \{0,\ldots, p\}$,
\begin{equation}\label{eq_thm_spectre_bands}
\# \left( \sigma(H_{2N+1})\cap \left[ E_{\mathrm{min}}^i, E_{\mathrm{max}}^i \right]  \right) = 2N+2  .
\end{equation}
\end{enumerate}
\end{theoreme}
\bigskip

\noindent Hence all the eigenvalues of $H_{2N+1}$ are in the band spectrum of $H$. In particular there is no eigenvalue of $H_{2N+1}$ embedded in the spectral gaps of $H$.

\noindent As the results of \cite{BL} for the periodic Airy-Schr\"odinger operator, Theorem \ref{thm_spectre} is not stated in the semiclassical limit, but for the semiclassical parameter $\cc$ larger than an explicit constant. As we will see in the proof of Lemma \ref{lem_variations_varphi} in Appendix \ref{sec_appendix_lemma_h}, the needed estimates become much more difficult to prove when $\cc$ is close to $c_p$ than when $\cc$ can be taken arbitrarily large.

\noindent For general results in the semiclassical limit for multiple wells with singularities at their minima, we refer to \cite{Y}. These results apply to $H_{2N+1}$. But, in our example, we are able to have more precise results on the counting of the eigenvalues and, as said before, our results are not only valid in the semiclassical limit but for values of the semiclassical parameter larger than a determined constant. This was made possible because (\ref{eq_schrodinger}) leads to an equation in $E$ which is solvable with classical special functions.

\noindent Theorem \ref{thm_spectre} also gives an example for general results such as \cite[Proposition 2.9]{HS1}. In our case, we obtain a precise counting result of the eigenvalues in the spectral bands of the corresponding periodic operator and the absence of eigenvalues embedded in the spectral gaps. To be more precise, the results of \cite{HS1} does not apply to $H_{2N+1}$ since the potential is singular at its minima and maxima points. But one can still see our results as an illustration of the general picture depicted in \cite{HS1}.



\subsection{The integrated density of states of the periodic Airy-Schr\"odinger operator}

The integrated density of states (IDS for short) of $H$ is the distribution function of its energy levels per unit volume. To define it properly, we can restrict the operator $H$ to finite length intervals and consider a thermodynamical limit. Since the definition of the domain of $H_{2N+1}$ includes Dirichlet boundary conditions at $\pm (2N+1)L_0$, one can take the following definition for the IDS associated to $H$.

\begin{defi}\label{def_ids}
The IDS associated to $H$ is the function from $\R$ to $\R_{+}$, $E\mapsto I(E)$, where $I(E)$, for $E\in \R$, is defined by 
\begin{equation}\label{eq_def_ids}
I(E)=\lim_{N\to +\infty} \frac{1}{2(2N+1)L_0} \# \{ \lambda \leq E\ ;\ \lambda \in \sigma(H_{2N+1}) \}.
\end{equation}
\end{defi}

\noindent For the existence of this limit and general properties of the IDS, we refer to the textbooks \cite{CL,Ki}.

\noindent To determine the expression of the IDS of $H$, one has to localize more precisely the eigenvalues of $H_{2N+1}$. To solve equation (\ref{eq_schrodinger}) we rescale it by setting $\E= \cc \frac{E}{V_0}$ and $\mathbf{V}_{2N+1}$, the multiplication operator by the rescaled potential $\mathbf{V}_{2N+1} : z\mapsto \sum_{k=-N}^N \mathbf{V}(z-2k)$ where 
$$ \mathbf{V}(z)=\left\lbrace \begin{array}{ccl}
                  |z| -1 & \mbox{ if } & z \in [-1,1] \\
		  0 & \mbox{ elsewhere} &
 \end{array}
 \right. .$$
We also set, for every $p\geq 0$, $ \E_{\mathrm{min}}^p =\cc \frac{E_{\mathrm{min}}^p}{V_0}$ and  $ \E_{\mathrm{max}}^p =\cc \frac{E_{\mathrm{max}}^p}{V_0}$. With $x=L_0z$ in (\ref{eq_schrodinger}), this equation is equivalent to 
\begin{align}
& \HH_{2N+1}\phi:=  -\frac{\mathrm{d}^2 }{\mathrm{d}z^2} \phi +\cc^3 \mathbf{V}_{2N+1}\phi=\cc^2 \E\phi  ,\quad \E\in \R,\label{eq_schrodinger_rescaled} \\
& \quad  \phi\in H^2(\R),\ \phi(-(2N+1))=\phi(2N+1). \nonumber
\end{align}
which defines the rescaled operator $\HH_{2N+1}$. One similarly defines the rescaled periodic Airy-Schr\"odinger operator $\HH$ whose spectral bands are the intervals $[\E_{\mathrm{min}}^p, \E_{\mathrm{max}}^p]$.

\noindent As $\mathbf{V}_{2N+1}$ is continuous and bounded, any solution $\phi$ of (\ref{eq_schrodinger_rescaled}) is continuously differentiable on $\R$.

\noindent In the sequel, we will restrict ourselves to energies $\E$ in the range of $\cc \mathbf{V}_{2N+1}$, the interval $[-\cc, 0]$.

\noindent To give the expression of the IDS of $\HH$ we need to introduce more notations. Let $\E\in[-\cc,0]$. We denote by $\U$ and $\V$ the functions defined, for every $x\in \R$, by 
\begin{equation}\label{def_U}
 \U(x)=v'(-\cc -\E)u(x)-u'(-\cc -\E)v(x)
\end{equation}
and
\begin{equation}\label{def_V}
\V(x)= u(-\cc -\E)v(x)-v(-\cc -\E)u(x).
\end{equation}
In particular, the Wronskian of $\U$ and $\V$ is equal to $1$. These functions $\U$ and $\V$ appear in the equations characterizing the spectral band edges of $\HH$ (\cite[(3.18)-(3.21)]{BL}). Indeed, the spectral band edges of $\HH$ are exactly the zeroes of $\U$, $\V$ and their derivatives.

\noindent We define the function 
\begin{equation}\label{def_argument}
\varphi\ :\ \left\lbrace \begin{array}{ccl}
                         \sigma(\HH) & \to & [0,\pi] \\
\E & \mapsto & \mathrm{Arg}((\U'\V+\U\V' + 2\ii\sqrt{-\U\U'\V\V'})(-\E) )
                         \end{array} \right.
\end{equation}
where for any complex number $a+\ii b$, $\mathrm{Arg}(a+\ii b)$ denotes its argument. We also denote by $\varphi$ the extension of the function $\varphi$ to $\R$ which maps any value outside of $\sigma(\HH)$ to $0$.

\noindent Also denote the integer part of a real number $x$ by $[x]$ and the characteristic function of an interval $[c,d]$ by $\mathbf{1}_{[c,d]}$. 

\begin{theoreme}\label{thm_ids}
Assume that $\cc \geq c_0$. For any $\E \in [-\cc,0]$, the integrated density of states associated to $\HH$ is given by 
\begin{equation}\label{eq_exp_ids}
\mathbf{I}(\E)=\frac12 p(\E) + \left\lbrace \begin{array}{ccl}  
\left( \frac12-\frac{1}{2\pi} \varphi(\E) \right)\cdot \mathbf{1}_{ [E_{\mathrm{min}}^{p(\E)},E_{\mathrm{max}}^{p(\E)}] }(\E) &\mathrm{if} & p(\E)\ \mathrm{is\ even}\\ 
\frac{1}{2\pi} \varphi(\E) \cdot \mathbf{1}_{ [E_{\mathrm{min}}^{p(\E)},E_{\mathrm{max}}^{p(\E)}] }(\E) &\mathrm{if} & p(\E)\ \mathrm{is\ odd}
\end{array} \right.
\end{equation}
where $p(\E)$ is the smallest integer such that $\E \leq \E_{\mathrm{max}}^{p(\E)}$ and is given by
\begin{equation}\label{eq_exp_pE}
p(\E)= \left[\frac{4}{3\pi} (\cc + \E)^{\frac32} \right].
\end{equation}
 \end{theoreme}

\noindent In particular, we remark that the function $\E\mapsto \mathbf{I}(\E)$ is continuous on the interval $[-\cc,0]$. In the proof of Proposition \ref{prop_variations_arg}, we give an explicit formula  for $\varphi$ in terms of the functions $\U$ and $\V$ (see (\ref{eq_arg_arctan1}), (\ref{eq_arg_arctan2}) and (\ref{eq_arg_arctan3})).

\bigskip 

\noindent The formula (\ref{eq_exp_ids}) is the analog, for the continuous periodic Airy-Schr\"odinger operator, of the formula of the IDS in \cite{Q} for a discrete periodic Schr\"odinger operator. 

\noindent The behavior of $\mathbf{I}$ is different than the one given by the results in \cite{K} in the Schr\"odinger case. Our result compares to the case $n=1$ and $l=1$ in \cite{K}, which leads to an IDS with an asymptotic in $E^{\frac12}$ for $E$ large, different from our case. Again, the singularity at the minima points of the potential leads to a different behavior than in the regular case. The same differences for the asymptotic behavior are to be found in more recent papers like \cite{PS} or \cite{I} where the potential (or periodic pertubation in \cite{I}) has regularity properties stronger than in our setting.

\noindent In the case of a singular potential, we refer again to \cite{Y}. In particular, the result of Theorem \ref{thm_ids} is compatible with the lower bounds and upper bounds found in \cite[Propositions 6.6 and 6.7]{Y}. Note that the results of \cite{Y} are stated in the semiclassical limit, hence for a ``large enough'' semiclassical parameter. No explicit lower bound for the semiclassical parameter gives the validity domain of \cite[Propositions 6.6 and 6.7]{Y}, compared to Theorem \ref{thm_ids} which states that the given formula is valid for every $\cc$ larger than the constant $c_0$. Note that $c_0$ is explicitely known as a zero of a classical function.

\noindent The results of \cite{I,K,PS} on the IDS are stated for general periodic potentials with some regularity assumption (in \cite{K} it is given through its Fourier coefficients). It leads to precise asymptotics in energy for the IDS. The results for multiple wells in \cite{Y} are given for a general $C^2$ potential which is $C^3$ in a neighborhood of some non-critical energy and leads to estimates on the IDS of the corresponding Schr\"odinger operator in the semiclassical limit. Here, since we look at a particular example of periodic operator, we have been able not only to find the asymptotics of the IDS as in \cite{I,K,PS} or an estimate as in \cite{Y}, but an explicit formula for it.








\section{The spectrum of $\HH_{2N+1}$}


\subsection{Characterization of the eigenvalues of $\HH_{2N+1}$}

\noindent We start by characterizing the real numbers $\E$ in $[-\cc,0]$ such that (\ref{eq_schrodinger_rescaled}) has a solution $\phi \in H^2(\R)$ which is not identically equal to zero and such that $\phi(-(2N+1))=\phi(2N+1)$.

\noindent From the canonical solutions $u$ and $v$ of the Airy equation, one defines the canonical pair of odd and even solutions of (\ref{eq_schrodinger_rescaled}) on the interval $[-1,1]$. The functions  defined for every $z\in [-1,1]$ by
\[ U_{\mathrm{even}}(z)=\U(\cc \mathbf{V}_{2N+1}(z)-\E)\]
and 
\[ V_{\mathrm{odd}}(z)=\mathrm{sign}(z)\V(\cc \mathbf{V}_{2N+1}(z)-\E)\]
form a basis of even and odd $C^1$ solutions of the equation (\ref{eq_schrodinger_rescaled}) on the interval $[-1,1]$. Indeed, one checks that $U_{\mathrm{even}}$ and $V_{\mathrm{odd}}$ are locally in the Sobolev space $\mathsf{H}^2(\R)$, thus are in $C^1(\R)$. Their wronskian satisfies 
$$\forall z\in [-1,1],\ (U_{\mathrm{even}}V_{\mathrm{odd}}^{'} - U_{\mathrm{even}}^{'}V_{\mathrm{odd}})(z)= \cc.$$

\noindent In any interval of the form $[2n-1,2n+1]$ for $n\in \{-N,\ldots, N\}$, a solution $\phi$ of (\ref{eq_schrodinger_rescaled}) writes
\begin{equation}
\forall z\in [2n-1,2n+1],\ \phi(z)=A_n U_{\mathrm{even}}(z-2n) + B_n V_{\mathrm{odd}}(z-2n) 
\end{equation}
where $A_n$ and $B_n$ are real numbers. Continuity of the solution $\phi$ and of its derivative at $2n+1$ for $n\in \{-N+1,\ldots , N-1 \}$ yields
\begin{align}
A_n \U(-\E) + B_n \V(-\E) & = A_{n+1} \U(-\E) - B_{n+1} \V(-\E) \label{eq_cont_milieu} \\
\cc \left(A_n \U'(-\E) + B_n \V' (-\E)\right) & =-\cc \left( A_{n+1} \U'(-\E) - B_{n+1} \V'(-\E) \right) \label{eq_deriv_milieu} 
\end{align}
The minus sign in front of $B_{n+1}$ in both (\ref{eq_cont_milieu}) and (\ref{eq_deriv_milieu}) comes from the oddness of $V_{\mathrm{odd}}$.

\noindent One deduces from  (\ref{eq_cont_milieu}) and (\ref{eq_deriv_milieu}) that for every $n\in \{ -N+1,\ldots , N-1 \}$,
\begin{equation}\label{eq_mat_transfer1}
\begin{pmatrix} A_{n+1} \\ B_{n+1} \end{pmatrix} =  \begin{pmatrix}
 (\U \V' +  \U'  \V)(-\E) & 2 (\V \V')(-\E) \\
 2 (\U \U')(-\E) &  (\U \V' +  \U'  \V)(-\E) 
\end{pmatrix} \begin{pmatrix} A_{n} \\ B_{n} \end{pmatrix}.
\end{equation}
using that the Wronskian of $\U$ and $\V$ is constant and equal to $1$. 

\noindent \textbf{Notations.} We set 
\begin{equation}\label{eq_def_a_b0_b1}
a:= (\U \V' +  \U'  \V)(-\E),\quad  b_0:=\sqrt{(\U \U')(-\E)}\ \mbox{ and }\ b_1:=\sqrt{(\V \V')(-\E)}.
\end{equation}
The squareroots are either positive real numbers or purely imaginary complex numbers with positive imaginary part depending on the signs of $\U \U'$ and $\V \V'$.

\noindent We introduce the transfer matrix which maps the solution of (\ref{eq_schrodinger_rescaled}) and its derivative on the interval $[2n-1,2n+1]$ to the solution and its derivative on the interval $[2n+1,2n+3]$:
\begin{equation}\label{eq_def_mat_transfer}
T_{\E} :=   \begin{pmatrix}
 a & 2 b_1^2 \\
 2 b_0^2 &  a 
\end{pmatrix}.
\end{equation}
In particular we have 
\begin{equation}\label{eq_rel_AN_A0}
\begin{pmatrix} A_{N} \\ B_{N} \end{pmatrix} = T_{\E}^{2N+1} \begin{pmatrix} A_{-N} \\ B_{-N} \end{pmatrix} . 
\end{equation}
Now we turn to the $C^1$ conditions at $-(2N+1)$ and $2N+1$. On the interval $[2N+1,+\infty)$, the potential $\mathbf{V}_M$ is identically zero, thus, since $\phi$ is in particular in $L^2(\R)$, one has
\begin{equation}\label{eq_cond_2N1}
\forall z \in [2N+1,+\infty),\ \phi(z)=\phi(2N+1) \ee^{-\lambda (z-(2N+1))} 
\end{equation}
with $\lambda$ a positive real number such that
$$-\lambda^2 =\cc^2  \E.$$
Without loss of generality one can choose $\phi$ such that $\phi(2N+1)=1$. Indeed, $\phi$ is not identically equal to zero and $(\phi(2N+1),\phi'(2N+1))=\phi(2N+1).(1,-\lambda)$. Hence by the Cauchy-Lipschitz theorem, $\phi(2N+1)\neq 0$. Then, since $\phi'(2N+1)=-\lambda$, one has
\begin{align}
A_{N} \U(-\E) + B_{N} \V(-\E) & = 1 \label{eq_cont_AN} \\
\cc \left(A_{N} \U'(-\E) + B_{N} \V' (-\E)\right) & =-\lambda .\label{eq_deriv_AN} 
\end{align}
Similarly, on the interval $(-\infty,-(2N+1)]$,  one has
\begin{equation}\label{eq_cond_0}
\forall z \in (-\infty,-(2N+1)] ,\ \phi(z)=\phi(-(2N+1)) \ee^{\lambda (z+2N+1)} 
\end{equation}
with the same $\lambda$ as in (\ref{eq_cond_2N1}). Since $\phi(-(2N+1))=\phi(2N+1)=1$, one has
\begin{align}
A_{-N} \U(-\E) - B_{-N} \V(-\E) & = 1 \label{eq_cont_A0} \\
\cc \left(A_{-N} \U'(-\E) - B_{-N} \V' (-\E)\right) & =\lambda .\label{eq_deriv_A0} 
\end{align}
Solving the system formed by the equations  (\ref{eq_cont_A0}) and (\ref{eq_deriv_A0}) one deduces that
\begin{equation}\label{eq_solve_A0_B0}
A_{-N} = \lambda  \frac{1}{\cc^2}  \V(-\E) +   \frac{1}{\cc} \V'(-\E) \ \mbox{ and }\ B_{-N}= \lambda \frac{1}{\cc^2} \U(-\E) +  \frac{1}{\cc} \U'(-\E). 
\end{equation}

\noindent Combining (\ref{eq_cont_AN}) and (\ref{eq_deriv_AN}),
\begin{equation}\label{eq_rel_AN}
\lambda \left( A_{N} \U(-\E) + B_{N} \V(-\E) \right) = - \cc \left(A_{N} \U'(-\E) + B_{N} \V' (-\E)\right).
\end{equation}
We also remark that, since $\lambda$ is positive, $\frac{\lambda}{\cc} = (-\E)^{\frac12}$. It leads to introduce the following function defined for $y\in [-\cc, 0]$ by:
\[
\Phi(y)= (y+\cc)^{\frac12} \left( A_{N} \U(y+\cc) + B_{N} \V(y+\cc) \right)  +A_{N} \U'(y+\cc) + B_{N} \V' (y+\cc)
\]
Then, $(\ref{eq_rel_AN})$ is equivalent to the equation in $\E$, $\Phi(-\cc-\E)=0$. The change of variables $y=-\cc-\E$ leads us to introduce for any $p\geq 0$,
$$Y_{\mathrm{min}}^p=-\cc - \E_{\mathrm{min}}^p \quad \mbox{and}\quad Y_{\mathrm{max}}^p=-\cc - \E_{\mathrm{max}}^p. $$
Remark that $\E \in [\E_{\mathrm{min}}^p,\E_{\mathrm{max}}^p ]$ if and only if  $y \in [Y_{\mathrm{max}}^p,Y_{\mathrm{min}}^p ]$. Hence, in the sequel, spectral bands or spectral gaps will refer indifferently to the intervals in the variable $\E$ or $y$.
\vskip 5mm

\section{Proof of Theorem \ref{thm_spectre}}

\subsection{Preliminary : the signs of $\U$, $\U'$, $\V$, $\V'$.}\label{sec_signs} From the results of \cite[Section 3.2]{BL}, the spectral band edges are exactly the zeroes of these four functions. Let us be more precise. Assume that $j\geq 1$ is an integer:
\begin{enumerate}
 \item The function $y\mapsto \U(y+\cc)$ vanishes at $Y_{\mathrm{max}}^{4j+2}$, is strictly negative on the interval $(Y_{\mathrm{max}}^{4j+2}, Y_{\mathrm{max}}^{4j})$, vanishes at $Y_{\mathrm{max}}^{4j}$ and is strictly positive on the interval $(Y_{\mathrm{max}}^{4j}, Y_{\mathrm{max}}^{4j-2})$.
 \item The function $y\mapsto \U'(y+\cc)$ vanishes at $Y_{\mathrm{min}}^{4j+2}$, is strictly negative on the interval $(Y_{\mathrm{min}}^{4j+2}, Y_{\mathrm{min}}^{4j})$, vanishes at $Y_{\mathrm{min}}^{4j}$ and is strictly positive on the interval $(Y_{\mathrm{min}}^{4j}, Y_{\mathrm{min}}^{4j-2})$.
 \item The function $y\mapsto \V(y+\cc)$ vanishes at $Y_{\mathrm{max}}^{4j+1}$, is strictly positive on the interval $(Y_{\mathrm{max}}^{4j+1}, Y_{\mathrm{max}}^{4j-1})$, vanishes at $Y_{\mathrm{max}}^{4j-1}$ and is strictly negative on the interval $(Y_{\mathrm{max}}^{4j-1}, Y_{\mathrm{max}}^{4j-3})$.
 \item The function $y\mapsto \V'(y+\cc)$ vanishes at $Y_{\mathrm{min}}^{4j+1}$, is strictly positive on the interval $(Y_{\mathrm{min}}^{4j+1}, Y_{\mathrm{min}}^{4j-1})$, vanishes at $Y_{\mathrm{min}}^{4j-1}$ and is strictly negative on the interval $(Y_{\mathrm{min}}^{4j-1}, Y_{\mathrm{min}}^{4j-3})$.
\end{enumerate}

{\footnotesize\begin{table}[H]
\begin{tikzpicture}
\tkzTabInit[lgt = 1.6, espcl = 1, deltacl = 0.5]{$y$ /1, band or gap /1 , $\U$ /1, $\U'$ /1, $\V$ /1, $\V'$ /1 } { $Y_{\mathrm{max}}^{4j+2}$, $Y_{\mathrm{min}}^{4j+2}$, $Y_{\mathrm{max}}^{4j+1}$, $Y_{\mathrm{min}}^{4j+1}$, $Y_{\mathrm{max}}^{4j}$ ,  $Y_{\mathrm{min}}^{4j}$, $Y_{\mathrm{max}}^{4j-1}$,  $Y_{\mathrm{min}}^{4j-1}$, $Y_{\mathrm{max}}^{4j-2}$,  $Y_{\mathrm{min}}^{4j-2}$, $Y_{\mathrm{max}}^{4j-3}$, $Y_{\mathrm{min}}^{4j-3}$}
\tkzTabLine{  , \text{band} ,   , \text{gap} ,  , \text{band} ,  , \text{gap}  ,   , \text{band} ,  , \text{gap}  ,  , \text{band} ,  , \text{gap}  , , \text{band} , , \text{gap},  , \text{band} , } 
\tkzTabLine{ z , - , , - , ,- ,  ,-  , z  , + , , +  ,  , + , , +  , z , - , , -, , -,  } 
\tkzTabLine{  , + , z , - , ,- ,  ,-  ,   , - , z , +  ,  , + , , +  ,  , + , z , - , ,- ,  } 
\tkzTabLine{  , - , , - ,z , + ,  , +  ,   , + , , +  , z , - ,  , -  ,  , - ,  , - , z , + ,   } 
\tkzTabLine{  , - , , - , ,- , z , +  ,   , + , , +  ,  , + , z , -  ,  , - , , -, , -, z } 
%
\end{tikzpicture}
\caption{Signs of $\U$, $\U'$, $\V$, $\V'$ on the interval $[Y_{\mathrm{max}}^{4j+2}, Y_{\mathrm{min}}^{4j-3}]$ for $j\geq 1$.}\label{tab_var_uv}

\end{table} }

\subsection{Expression of $\Phi$}

We have to compute the coefficients $A_N$ and $B_N$ to get the explicit expression of $\Phi$. Thus we have to compute the matrix $T_{\E}^{2N+1}$. It suffices to diagonalize $T_{\E}$. For this, it is easier to separate the cases when $\E$ is in a spectral gap of $\HH$, $\E$ is in an even spectral band (for an index $p$ even) of $\HH$ and when $\E$ is in an odd spectral band (for an index $p$ odd) of $\HH$.

\subsubsection{In the spectral gaps.} For $\E$ in a spectral gap, we have $(\U \U')(-\E) >0$ and $(\V \V')(-\E)>0$. Thus, the eigenvalues of $T_{\E}$ are 
\begin{equation}\label{eq_evs_TE_gaps}
 a+2b_0 b_1 \mbox{ and } a-2b_0 b_1
\end{equation}
and the associated eigenvectors are
\begin{equation}\label{eq_vps_TE_gaps}
\begin{pmatrix}
 b_1\\
 b_0
\end{pmatrix} \quad \mbox{and}\quad \begin{pmatrix}
 -b_1\\
 b_0
\end{pmatrix}.
\end{equation}
\noindent We set 
\begin{equation}\label{eq_def_b}
 b:= 2 \sqrt{(\U \U' \V \V')(-\E)} = 2b_0 b_1.
\end{equation}
We note that, for any $\E$ in a spectral gap, $2 \sqrt{(\U \U' \V \V')(-\E)} <  |(\U \V' +  \U'  \V)(-\E)|$. Indeed, if $2 \sqrt{(\U \U' \V \V')(-\E)} = |(\U \V' +  \U'  \V)(-\E)|$ then $((\U \V')(-\E) - (\U' \V)(-\E))^2=0$ but $ \U \V' -  \U'  \V$ is constant and equal to $1$ since the Wronskian of $u$ and $v$ is equal to $1$. Thus, 
\begin{equation}\label{eq_ineg_a_b}
 0< b < |a|.
\end{equation}
Then
$$T_{\E}^{2N+1}=\frac{1}{2^{2N+1}} \left( \begin{array}{cc}
                        b_1 & -b_1\\
 b_0 &  b_0 
                        \end{array} \right)  \left( \begin{array}{cc}
                    (a+b)^{2N+1}     & 0 \\
0 &   (a-b)^{2N+1} \end{array} \right)  \left( \begin{array}{cc}
                        \frac{1}{b_1} & \frac{1}{ b_0 }\\
 -\frac{1}{b_1}&  \frac{1}{ b_0 } 
                        \end{array} \right).  $$
This gives the expressions of the coefficients $A_N$ and $B_N$ and thus the expression of $\Phi(y)$ for $y$ such that $-\E$ is in a gap of $\HH$ and $y\in [-\cc,0]$:
\begin{align}\label{eq_expr_Phi_gaps}
\Phi(y) & = \frac{1}{2^{2N}} \left[ ((\U' + (\cdot )^{\frac12}\U )(\V' + (\cdot )^{\frac12}\V ))(y+\cc)\times ((a+b)^{2N+1}+(a-b)^{2N+1}) \right. \nonumber \\
& \left. + \left( \frac{b_1}{b_0} (\U' + (\cdot)^{\frac12}\U )^2 + \frac{b_0}{b_1}  (\V' + (\cdot)^{\frac12}\V )^2  \right)(y+\cc)\times  ((a+b)^{2N+1}-(a-b)^{2N+1})\right]
\end{align}


\subsubsection{In the interior of even spectral bands.} For $\E \in (\E_{\mathrm{min}}^{2j},\E_{\mathrm{max}}^{2j} )$ , we have $(\U \U')(-\E) <0$ and $(\V \V')(-\E)>0$. Thus, the eigenvalues of $T_{\E}$ are 
\begin{equation}\label{eq_evs_TE_bands_2j}
 a + 2\ii \sqrt{-(\U \U' \V \V')(-\E)} \mbox{ and } a - 2\ii \sqrt{-(\U \U' \V \V')(-\E)},
\end{equation}
and the associated eigenvectors are
\begin{equation}\label{eq_vps_TE_bands_2j}
\begin{pmatrix}
b_1\\
b_0
\end{pmatrix} \quad \mbox{and}\quad \begin{pmatrix}
b_1\\
-b_0
\end{pmatrix}
\end{equation}
with $b_0 = \ii \sqrt{-(\U \U' )(-\E)}$ in this case.
Assume that $y$ is such that $-\E$ is in an even spectral band of $\HH$ and $y\in [-\cc,0]$. 
We slightly change the previous notation of $b$ in order to highlight the complex number $\ii$ in the following expressions and we set 
\begin{equation}\label{eq_def_b_complex}
b(y+\cc):= 2 \sqrt{-(\U \U' \V \V')(y+\cc)} .
\end{equation}
We also set
\begin{equation}\label{eq_def_alpha_beta}
\alpha(y+\cc):=\left( \sqrt{\frac{\V\V'}{-\U \U'}}(\U' + (\cdot)^{\frac12}\U )^2 \right)(y+\cc)  \mbox{ and } \beta(y+\cc):=\left( \sqrt{\frac{-\U \U'}{\V\V'}}  (\V' + (\cdot)^{\frac12}\V )^2  \right)(y+\cc)  
\end{equation}
Then, the expression of $\Phi(y)$ is:
\begin{align}
\Phi(y) & = \frac{1}{2^{2N+1}} \left[ 2((\U' + (\cdot )^{\frac12}\U )(\V' + (\cdot )^{\frac12}\V ))(y+\cc)\times ((a+\ii b)^{2N+1} + (a-\ii b)^{2N+1})(y+\cc)  \right. \nonumber \\
& \left. \ \ \ + \left( \ii \alpha - \ii \beta \right)(y+\cc) \times ((a+\ii b)^{2N+1} - (a-\ii b)^{2N+1})(y+\cc) \right] \nonumber \\
& = \frac{1}{2^{2N+1}} ((a+\ii b)^{2N+1} + (a-\ii b)^{2N+1})(y+\cc) \left[ 2((\U' + (\cdot )^{\frac12}\U )(\V' + (\cdot )^{\frac12}\V ))(y+\cc)\right. \nonumber \\
& \left. \ \ \ + \left( \alpha - \beta  \right)(y+\cc) \times \tan((2N+1) \mathrm{Arg}(a+\ii b)(y+\cc))\right] \nonumber \\
& = \frac{1}{2^{2N+1}} ((a+\ii b)^{2N+1} + (a-\ii b)^{2N+1})(y+\cc)   \times \left( \alpha + \beta  \right)(y+\cc)  \nonumber \\
& \times \left[ \left( \frac{2(\U' + (\cdot )^{\frac12}\U )(\V' + (\cdot )^{\frac12}\V )}{  \alpha + \beta  } \right)(y+\cc) + \left( \frac{\alpha - \beta}{ \alpha + \beta }   \right)(y+\cc) \times \tan((2N+1) \mathrm{Arg}(a+\ii b)(y+\cc))\right]. \label{eq_expr_Phi_bands_2j_1}
\end{align}
We also used that
\begin{equation}
 \sqrt{ \left( 2(\U' + (\cdot )^{\frac12}\U )(\V' + (\cdot )^{\frac12}\V ) \right)^2 + \left( \alpha - \beta \right)^2  } = \alpha + \beta.
\end{equation}
Let us introduce the function $k : \sigma(\HH)\cap [-\cc,0] \to \R$ define by
\begin{equation}\label{def_function_k}
\forall y \in  \sigma(\HH)\cap [-\cc,0],\ k(y)=\left( \sqrt{-\frac{\V\V'}{\U\U'}}\times \frac{\U' + (\cdot)^{\frac12}\U }{\V' + (\cdot)^{\frac12}\V} \right) (y+\cc). 
\end{equation}
Then, (\ref{eq_expr_Phi_bands_2j_1}) rewrites, for $y$ such that $-\E$ is in an even spectral band of $\HH$ and $y\in [-\cc,0]$ and since $\beta(y+\cc)>0$:
\begin{align}
\Phi(y) & = \frac{1}{2^{2N+1}} ((a+\ii b)^{2N+1} + (a-\ii b)^{2N+1})(y+\cc)\times    \left( \alpha + \beta  \right)(y+\cc)  \nonumber \\
& \times \left[ \frac{2k(y)}{ 1+(k(y))^2} +  \frac{ (k(y))^2-1 }{ (k(y))^2 +1 }   \times \tan((2N+1) \mathrm{Arg}(a+\ii b)(y+\cc))\right]. \label{eq_expr_Phi_bands_2j}
\end{align}

\subsubsection{In the interior of odd spectral bands.} For $\E \in (\E_{\mathrm{min}}^{2j+1},\E_{\mathrm{max}}^{2j+1} )$ , we have $(\U \U')(-\E) >0$ and $(\V \V')(-\E)<0$. The eigenvalues of $T_{\E}$ are again
\begin{equation}\label{eq_evs_TE_bands_2j1}
  a + \ii b \mbox{ and } a - \ii b
\end{equation}
and the associated eigenvectors are
\begin{equation}\label{eq_vps_TE_bands_2j1}
\begin{pmatrix}
b_1\\
b_0  
\end{pmatrix} \quad \mbox{and}\quad \begin{pmatrix}
-b_1\\
 b_0
\end{pmatrix}
\end{equation}
with $b_1 = \ii \sqrt{-(\V \V' )(-\E)}$ in this case.

The expression of $\Phi(y)$ is, for $y$ such that $-\E$ is in an odd spectral band of $\HH$ and $y\in [-\cc,0]$:
\begin{align}
\Phi(y) & = \frac{1}{2^{2N+1}} ((a+\ii b)^{2N+1} + (a-\ii b)^{2N+1})(y+\cc) \times \left( \alpha + \beta  \right)(y+\cc)  \nonumber \\
& \times \left[ \frac{2\tilde{k}(y)}{ 1+(\tilde{k}(y))^2} + \frac{ (\tilde{k}(y))^2-1 }{ (\tilde{k}(y))^2 +1 }   \times \tan((2N+1) \mathrm{Arg}(a+\ii b)(y+\cc))\right].
\label{eq_expr_Phi_bands_2j1}
\end{align}
setting for $y \in  \sigma(\HH)\cap [-\cc,0]$, $\tilde{k}(y)=\frac{1}{k(y)}$, since $k(y)$ does not vanish in the interior of the odd spectral bands.

\subsection{Counting the zeroes of $\Phi$ in $[-\cc,0]$.}\label{sec_proof_counting} In this section, we prove that $\Phi$ does not vanish in the spectral gaps and has $2N+2$ zeroes in each spectral band contained in $[-\cc,0]$.

\subsubsection{In the spectral gaps.}\label{sec_no_evs_gaps} Let $j\geq 1$ be an integer. From the signs of $\U$, $\U'$, $\V$ and $\V'$ given in Section \ref{sec_signs}, the function $y\mapsto \U'(y+\cc)+(y+\cc)^{\frac12}\U(y+\cc)$ is strictly positive in the spectral gap $(Y_{\mathrm{min}}^{4j+3},Y_{\mathrm{max}}^{4j+2})$, strictly negative in the spectral gaps $(Y_{\mathrm{min}}^{4j+2},Y_{\mathrm{max}}^{4j+1})$ and $(Y_{\mathrm{min}}^{4j+1},Y_{\mathrm{max}}^{4j})$ and strictly positive in the spectral gap $(Y_{\mathrm{min}}^{4j},Y_{\mathrm{max}}^{4j-1})$.

\noindent The function $y\mapsto \V'(y+\cc)+(y+\cc)^{\frac12}\V(y+\cc)$ is strictly negative in the spectral gaps $(Y_{\mathrm{min}}^{4j+3},Y_{\mathrm{max}}^{4j+2})$ and $(Y_{\mathrm{min}}^{4j+2},Y_{\mathrm{max}}^{4j+1})$, and strictly positive in the spectral gaps $(Y_{\mathrm{min}}^{4j+1},Y_{\mathrm{max}}^{4j})$ and $(Y_{\mathrm{min}}^{4j},Y_{\mathrm{max}}^{4j-1})$.
\vskip 3mm

\noindent \emph{First case.} Assume that we are in the spectral gap $(Y_{\mathrm{min}}^{4j+2},Y_{\mathrm{max}}^{4j+1})$ or in the spectral gap  $(Y_{\mathrm{min}}^{4j},Y_{\mathrm{max}}^{4j-1})$. Then the product $( \U'(y+\cc)+(y+\cc)^{\frac12}\U(y+\cc))(\V'(y+\cc)+(y+\cc)^{\frac12}\V(y+\cc))$ is strictly positive and the functions $y\mapsto (\U\V')(y+\cc)$ and $y\mapsto (\U'\V)(y+\cc)$ are also strictly positive. Thus, $a>0$, $b>0$, $a+b>0$ and, using (\ref{eq_ineg_a_b}), $a-b>0$. Moreover, since $b>0$, we always have $a+b>a-b$. According to the expression (\ref{eq_expr_Phi_gaps}), $\Phi(y)>0$ in the spectral gaps $(Y_{\mathrm{min}}^{4j+2},Y_{\mathrm{max}}^{4j+1})$ and $(Y_{\mathrm{min}}^{4j},Y_{\mathrm{max}}^{4j-1})$ and in particular does not vanish in these intervals. 
\vskip 3mm

\noindent \emph{Second case.} Assume that we are in the spectral gap $(Y_{\mathrm{min}}^{4j+3},Y_{\mathrm{max}}^{4j+2})$ or in the spectral gap $(Y_{\mathrm{min}}^{4j+1},Y_{\mathrm{max}}^{4j})$. Then the product $( \U'(y+\cc)+(y+\cc)^{\frac12}\U(y+\cc))(\V'(y+\cc)+(y+\cc)^{\frac12}\V(y+\cc))$ is strictly negative and the functions $y\mapsto (\U\V')(y+\cc)$ and $y\mapsto (\U'\V)(y+\cc)$ are also strictly negative. Thus, $a<0$, $b>0$, and using (\ref{eq_ineg_a_b}), $a-b<0$ and $a+b<0$. Since $2N+1$ is odd, $(a+b)^{2N+1}<0$, $(a-b)^{2N+1}<0$ and $(a+b)^{2N+1}+(a-b)^{2N+1}<0$. The first term in the expression of $\Phi(y)$ in (\ref{eq_expr_Phi_gaps}) is a product of two strictly negative real numbers and is strictly positive. We also have $0>a+b>a-b$, hence, again by oddness of $2N+1$, $(a+b)^{2N+1}>(a-b)^{2N+1}$ and the second term in  (\ref{eq_expr_Phi_gaps}) is a product of two strictly positive real numbers and is also strictly positive. Thus, $\Phi(y)>0$ in the spectral gaps $(Y_{\mathrm{min}}^{4j+3},Y_{\mathrm{max}}^{4j+2})$ and $(Y_{\mathrm{min}}^{4j+1},Y_{\mathrm{max}}^{4j})$ and does not vanish in them.
\vskip 3mm

\noindent From these two cases, we deduce that $H_{2N+1}$ does not have any eigenvalue in the spectral gaps of $H$, as long as we look at the spectral gaps in the interval $[-V_0,0]$. This proves the first point of Theorem \ref{thm_spectre}.

\vskip 3mm

\noindent Remark that in the second case the signs depends on the oddness of $2N+1$. They are slight changes to do in the even case and we present them in Appendix \ref{sec_appendix_M_even}.

\subsubsection{In the spectral bands.} Remark that from the expressions (\ref{eq_expr_Phi_bands_2j}) and (\ref{eq_expr_Phi_bands_2j1}), similar arguments will give the number of zeroes of $\Phi$ in the even and odd spectral bands. Thus we will focus on the case of the even spectral bands and explain the differences with this case in the case of odd spectral bands. 
\vskip 3mm

\noindent Using the expressions of $\Phi(y)$ given by (\ref{eq_expr_Phi_bands_2j}) and  (\ref{eq_expr_Phi_bands_2j1}), in order to count the zeroes of $\Phi$ in the spectral bands, we first have to study the properties of the function 
\begin{equation}\label{def_tilde_varphi}
\tilde{\varphi}\ :\ \begin{array}{ccl}
                   -\cc - \sigma(\HH) & \to & \R \\
 y& \mapsto & \mathrm{Arg}(a+\ii b)(y+\cc).
                   \end{array}
\end{equation}
We prove the following result.

\begin{prop}\label{prop_variations_arg}
 Let $p\geq 0$ and $\cc\geq c_p$.  Then, for every $l\in \{ 0,\ldots , p\}$, 
 \begin{enumerate}
 \item if $l=2j$ is even, the function $\tilde{\varphi}$ is a strictly decreasing homeomorphism from $[Y_{\mathrm{max}}^{2j},Y_{\mathrm{min}}^{2j} ]$ to $[0,\pi]$,
 \item  if $l=2j+1$ is odd, the function $\tilde{\varphi}$ is a strictly increasing homeomorphism from $[Y_{\mathrm{max}}^{2j+1},Y_{\mathrm{min}}^{2j+1} ]$ to $[0,\pi]$.
 \end{enumerate}
\end{prop}

\noindent Before proving Proposition \ref{prop_variations_arg} we give the following lemma.

\begin{lemma}\label{lem_variations_varphi}
 Let $p\geq 0$,  $\cc\geq c_p$ and $h\ : [-\cc,0] \to \R$ given by :
 \begin{equation}\label{def_h}
 \forall y\in [-\cc,0],\ h(y)=-(\U\U')(y+\cc) - y (\V\V') (y+\cc) + (\U'\V')(y+\cc) +(y+\cc)(\U\V)(y+\cc).
 \end{equation}
 Then, for every $l\in \{ 0,\ldots , p\}$, 
 \begin{enumerate}
 \item if $l=2j$ is even,  for every $y\in [Y_{\mathrm{max}}^{2j},Y_{\mathrm{min}}^{2j} ]$, $h(y)>0$,
 \item  if $l=2j+1$ is odd,  for every $y\in [Y_{\mathrm{max}}^{2j+1},Y_{\mathrm{min}}^{2j+1} ]$, $h(y)<0$. 
 \end{enumerate}
\end{lemma}
\noindent The proof of this lemma is quite technical and we postpone it until Appendix \ref{sec_appendix_lemma_h}.
\bigskip

\begin{demo} (of Proposition  \ref{prop_variations_arg}).

\noindent \emph{Step 1.} Let us start by computing the derivatives of the functions $y\mapsto \U(y+\cc)$, $y\mapsto \U'(y+\cc)$, $y\mapsto \V(y+\cc)$ and $y\mapsto \V'(y+\cc)$. Since, for every $y\in \R$, $\U(y+\cc)=v'(y)u(y+\cc)-u'(y)v(y+\cc)$ and since we have similar expressions for the three others functions, and because $u$ and $v$ are solutions of the Airy equation :
\begin{align}
\frac{\dd}{\dd y}  \U(y+\cc) & = -y\V(y+\cc) + \U'(y+\cc) \label{eq_derivee_U} \\
\frac{\dd}{\dd y}  \U'(y+\cc) & = -y\V'(y+\cc) + (y+\cc)\U(y+\cc) \label{eq_derivee_Up} \\
\frac{\dd}{\dd y}  \V(y+\cc) & = -\U(y+\cc) + \V'(y+\cc) \label{eq_derivee_V} \\
\frac{\dd}{\dd y}  \V'(y+\cc) & = -\U'(y+\cc) + (y+\cc)\V(y+\cc) \label{eq_derivee_Vp}
\end{align}
where the $'$ denote the derivation with respect to $x$ in the expressions (\ref{def_U}) and (\ref{def_V}). Using (\ref{eq_derivee_U}), (\ref{eq_derivee_Up}), (\ref{eq_derivee_V}) and (\ref{eq_derivee_Vp}), one gets :
\begin{equation}\label{eq_derivee_UUpVVp}
\frac{\dd}{\dd y} (\U \U' \V \V')(y+\cc) = (\U \V' + \V \U')(y+\cc)\times h(y).  
\end{equation}
Let $g\ :\ [-\cc,0] \to \R$ given by
\begin{equation}\label{def_g}
\forall y \in [-\cc,0],\ g(y)=  (\U \V' + \V \U')(y+\cc).
\end{equation}
Then, using again (\ref{eq_derivee_U}), (\ref{eq_derivee_Up}), (\ref{eq_derivee_V}) and (\ref{eq_derivee_Vp}), we have 
\begin{equation}\label{eq_derivee_g}
\forall y \in [-\cc,0],\ \frac{\dd}{\dd y} g(y) = 2h(y)
\end{equation}
and Lemma \ref{lem_variations_varphi} gives the variations of $g$ on each spectral band $[Y_{\mathrm{max}}^{l},Y_{\mathrm{min}}^{l} ]$. On each even spectral band ($l=2j$), $g$ is strictly increasing from $g(Y_{\mathrm{max}}^{2j})=(\V \U')(Y_{\mathrm{max}}^{2j})<0$ to $g(Y_{\mathrm{min}}^{2j})=(\U \V')(Y_{\mathrm{min}}^{2j})>0$, hence it vanishes at a unique point $Y_{0}^{2j} \in (Y_{\mathrm{max}}^{2j},Y_{\mathrm{min}}^{2j})$. Similarly, on each odd spectral band, $g$ is strictly decreasing and vanishes at a unique point $Y_{0}^{2j+1} \in (Y_{\mathrm{max}}^{2j+1},Y_{\mathrm{min}}^{2j+1})$. 
\bigskip 

\noindent \emph{Step 2.} We compute an explicit expression of the function $\tilde{\varphi}$ in terms of the functions $\U$, $\U'$, $\V$, $\V'$. Assume that $y \in [Y_{\mathrm{max}}^{l},Y_{\mathrm{min}}^{l} ]$ for some $l\in \{0,\ldots , p\}$.  Then,
\begin{align*}
|(a+\ii b)(y+\cc)| & = | (\U \V' + \U' \V + 2 \ii \sqrt{-\U \U' \V \V' } )(y+\cc)| \\
 & = \left( ((\U \V' + \U' \V)(y+\cc))^2 - 4(\U \U' \V \V' )(y+\cc) \right)^{\frac12} \\
& = | (\U \V' - \U' \V)(y+\cc) | = 1
\end{align*}
since the Wronskian of $\U$ and $\V$ is equal to $1$. Moreover, since $ 2 \sqrt{-\U \U' \V \V' }(y+\cc) \geq 0$, $ \mathrm{Arg}(a+\ii b)(y+\cc) \in [0,\pi]$. If $l=2j$ is even, $g(y)< 0$ on $[Y_{\mathrm{max}}^{2j},Y_{0}^{2j})$ and 
\begin{equation}\label{eq_arg_arctan1}
\forall y \in  [Y_{\mathrm{max}}^{2j},Y_{0}^{2j}),\ \tilde{\varphi}(y) = \pi + \mathrm{Arctan}\left( \frac{b}{a} \right)(y+\cc).
\end{equation}
On $(Y_{0}^{2j},Y_{\mathrm{min}}^{2j}]$, $g(y)>0$ and 
\begin{equation}\label{eq_arg_arctan2}
\forall y \in  (Y_{0}^{2j},Y_{\mathrm{min}}^{2j}],\ \tilde{\varphi}(y) = \mathrm{Arctan}\left( \frac{b}{a } \right)(y+\cc).
\end{equation}
Since $(a+\ii b)(Y_{0}^{2j}+\cc) = \ii b(Y_{0}^{2j}+\cc))$, $\mathrm{Arg}(a+\ii b)(Y_{0}^{2j}+\cc) = \frac{\pi}{2}$ and the function $\tilde{\varphi}$ is continuous on $[Y_{\mathrm{max}}^{2j},Y_{\mathrm{min}}^{2j}]$. 

\noindent If $l=2j+1$ is odd, similarly,
\begin{equation}\label{eq_arg_arctan3}
\tilde{\varphi}(y) = \left\lbrace \begin{array}{ccl}
  \mathrm{Arctan}\left( \frac{b}{a } \right)(y+\cc)& \mathrm{if} & y \in  [Y_{\mathrm{max}}^{2j+1},Y_{0}^{2j+1}) \\
\pi + \mathrm{Arctan}\left( \frac{b }{a} \right)(y+\cc) & \mathrm{if} & y \in  (Y_{0}^{2j+1},Y_{\mathrm{min}}^{2j+1}]
\end{array} \right. 
\end{equation}
and again, since,  $\mathrm{Arg}(a+\ii b)(Y_{0}^{2j+1}+\cc) = \frac{\pi}{2}$, the function $\tilde{\varphi}$ is continuous on $[Y_{\mathrm{max}}^{2j+1},Y_{\mathrm{min}}^{2j+1}]$. 
\bigskip 

\noindent \emph{Step 3.} It remains to prove the monotonicity assertions to prove Proposition \ref{prop_variations_arg}.  For every $y\in [Y_{\mathrm{max}}^{l},Y_{\mathrm{min}}^{l} ]$, let
$$f(y) = \left( \frac{2\sqrt{-\U\U'\V\V'}}{\U \V' + \U' \V } \right)(y+\cc).$$
Then, using (\ref{eq_derivee_UUpVVp}) and (\ref{eq_derivee_g}), 
\begin{align*}
\frac{\dd}{\dd y}  f(y) & = \frac{2 \frac{\frac{\dd}{\dd y}(-\U\U'\V\V')(y+\cc)  }{2\sqrt{-\U\U'\V\V'}} g(y) -2\sqrt{-\U\U'\V\V'}(y+\cc) \frac{\dd}{\dd y}g(y)}{(g(y))^2 }\\
& = \frac{h(y)\left( 4(\U\U'\V\V')(y+\cc) - (g(y))^2\right)}{(g(y))^2 \sqrt{-\U\U'\V\V'}(y+\cc)} \\
& = \frac{h(y)\left( -(\U\V'-\V \U')(y+\cc) \right)^2}{(g(y))^2 \sqrt{-\U\U'\V\V'}(y+\cc)} \\
& = \frac{-h(y)}{(g(y))^2 \sqrt{-\U\U'\V\V'}(y+\cc)}.
\end{align*}
since the Wronskian of $\U$ and $\V$ is equal to $1$. Using Lemma \ref{lem_variations_varphi}, $\frac{\dd}{\dd y}  f$ is strictly negative on each even spectral band and is strictly positive on each odd spectral band. Hence, $f$ is strictly decreasing on each even spectral band and by (\ref{eq_arg_arctan1}) and (\ref{eq_arg_arctan2}), $\tilde{\varphi}$ is strictly decreasing from $[Y_{\mathrm{max}}^{2j},Y_{\mathrm{min}}^{2j} ]$ to ``$[\pi,0]$''. Since it is continuous, point (1) of Proposition \ref{prop_variations_arg} follows. Similarly, $f$ is strictly increasing on each odd spectral band and by (\ref{eq_arg_arctan3}) and continuity proven at Step 2, point (2) of Proposition \ref{prop_variations_arg} follows.
\end{demo}

\noindent \textbf{Notation.} We set, for every $m\in \{1,\ldots , 2N\}$, 
\begin{equation}\label{eq_def_ym}
y_m^{\pm} =\tilde{\varphi}^{-1}\left( \frac{(2m\pm 1)\pi}{2(2N+1)} \right). 
\end{equation}

\begin{demo} (of Theorem \ref{thm_spectre}). We count the number of zeroes of $\Phi$ in an even spectral band $[Y_{\mathrm{max}}^{2j},Y_{\mathrm{min}}^{2j}]$, $2j\in \{0,\ldots, p\}$. The case of an odd spectral band is similar and we give the necessary changes to the proof of the even case at the end of this proof. 
\bigskip

\noindent \emph{Step 1.} By Proposition \ref{prop_variations_arg}, $\tilde{\varphi}$ is a strictly decreasing homeomorphism from $[Y_{\mathrm{max}}^{2j},Y_{\mathrm{min}}^{2j}]$ to $[0,\pi]$ with $\tilde{\varphi}(Y_{\mathrm{max}}^{2j})=\pi$ and $\tilde{\varphi}(Y_{\mathrm{min}}^{2j})=0$. Hence, the function $y\mapsto \tan((2N+1)\tilde{\varphi}(y))$
\begin{enumerate}
 \item vanishes at $Y_{\mathrm{max}}^{2j}$ ; 
\item  is continuous, strictly decreasing on $\left[Y_{\mathrm{max}}^{2j},y_{2N}^{+} \right)$ and tends to $-\infty$ when $y$ tends to $\left(y_{2N}^{+}\right)^{-}$ ; 
\item is continuous and strictly decreasing from $+\infty$ to $-\infty$ on each interval \\ $\left(y_m^+,y_m^- \right)$ for $m\in \{1,\ldots , 2N\}$ ; 
\item  is continuous, strictly decreasing on $( y_1^-, Y_{\mathrm{min}}^{2j}] $ and tends to $+\infty$ when $y$ tends to $\left(y_1^-\right)^{+}$ ;
\item vanishes at $Y_{\mathrm{min}}^{2j}$.
\end{enumerate}
By the intermediate value theorem, the function $y\mapsto \tan((2N+1)\tilde{\varphi}(y))$ has exactly $2N+2$ zeroes in the interval $[Y_{\mathrm{max}}^{2j},Y_{\mathrm{min}}^{2j}]$.
\bigskip

\noindent \emph{Step 2.} The function $k$ defined at (\ref{def_function_k}) is strictly increasing from $-\infty$ to $+\infty$ on the interval $[Y_{\mathrm{max}}^{2j},Y_{\mathrm{min}}^{2j}]$. 

\noindent The function $x\mapsto \frac{x^2 -1}{x^2 +1}$ is continous on $\R$, even and it is strictly increasing from $-1$ to $1$ on $[0,+\infty)$. Hence, $y\mapsto \frac{(k(y))^2 -1}{(k(y))^2 +1}$ is strictly positive on $[Y_{\mathrm{max}}^{2j},k^{-1}(-1))$, strictly negative on $(k^{-1}(-1), k^{-1}(1))$, strictly positive on $(k^{-1}(1),Y_{\mathrm{min}}^{2j} ]$ and vanishes only at the points $k^{-1}(-1)$ and $k^{-1}(1)$. It is also strictly decreasing from $1$ to $-1$ on $[Y_{\mathrm{max}}^{2j},k^{-1}(0)]$ and strictly increasing from $-1$ to $1$ on $[k^{-1}(0),Y_{\mathrm{min}}^{2j} ]$. Let $m_{\pm}$ the unique integer such that $k^{-1}(\pm 1) \in  (y_{m_{\pm}}^{+},y_{m_{\pm}}^{-} )$. Then, the function  $y\mapsto  \frac{(k(y))^2 -1}{(k(y))^2 +1} \tan((2N+1)\tilde{\varphi}(y))$
\begin{enumerate}
 \item vanishes at $Y_{\mathrm{max}}^{2j}$ ; 
\item  is continuous, strictly decreasing on $\left[Y_{\mathrm{max}}^{2j},y_{2N}^+ \right)$ and tends to $-\infty$ when $y$ tends to $\left(y_{2N}^+\right)^{-}$ ; 
\item is continuous and strictly decreasing from $+\infty$ to $-\infty$ on each interval \\ $\left(y_m^+,y_m^- \right)$ for $m\in \{1,\ldots , m_{-}-1\}$ ;
\item is continuous on $(y_{m_{-}}^+,y_{m_{-}}^- )$, strictly decreasing from $+\infty$ to $0$ on $(y_{m_{-}}^+, k^{-1}(-1)]$, negative on $\left[k^{-1}(-1), \tilde{\varphi}^{-1}\left(\frac{m_{-}\pi}{2N+1}\right)\right]$ and strictly increasing from $0$ to $+\infty$ on 
\\ $\left[\tilde{\varphi}^{-1}\left(\frac{m_{-}\pi}{2N+1}\right), y_{m_{-}}^- \right)$, if $k^{-1}(-1) <  \tilde{\varphi}^{-1}\left(\frac{m_{-}\pi}{2N+1}\right)$ (and the same variations if
\\ $k^{-1}(-1) >  \tilde{\varphi}^{-1}\left(\frac{m_{-}\pi}{2N+1}\right)$, exchanging these two real numbers in the previous intervals bounds) ;
\item is continuous and strictly increasing from $-\infty$ to $+\infty$ on each interval $\left(y_m^+,y_m^- \right)$ for $m\in \{m_{-}+1,\ldots , m_{+}-1\}$ ;
\item  is continuous on $(y_{m_{+}}^+ ,y_{m_{+}}^-)$, strictly increasing from $-\infty$ to $0$ on $(y_{m_{+}}^+, k^{-1}(1)]$, positive on $\left[k^{-1}(1), \tilde{\varphi}^{-1}\left(\frac{m_{+}\pi}{2N+1}\right)\right]$  and strictly decreasing from $0$ to $-\infty$ on 
\\ $\left[  \tilde{\varphi}^{-1}\left(\frac{m_{+}\pi}{2N+1}\right), y_{m_{+}}^- \right)$, if $k^{-1}(1) <  \tilde{\varphi}^{-1}\left(\frac{m_{+}\pi}{2N+1}\right)$ (and the same variations if $k^{-1}(1) >  \tilde{\varphi}^{-1}\left(\frac{m_{+}\pi}{2N+1}\right)$, exchanging these two real numbers in the previous intervals bounds)  ;
\item is continuous and strictly decreasing from $+\infty$ to $-\infty$ on each interval $\left(y_m^+,y_m^- \right)$ for $m\in \{m_{+}+1,\ldots , 2N\}$ ;
\item  is continuous, strictly decreasing on $( y_1^-, Y_{\mathrm{min}}^{2j}] $ and tends to $+\infty$ when $y$ tends to $\left(y_1^-\right)^{+}$ ;
\item vanishes at $Y_{\mathrm{min}}^{2j}$.
\end{enumerate}
Since $k^{-1}(-1) \neq  \tilde{\varphi}^{-1}\left(\frac{m_{-}\pi}{2N+1}\right)$ and $k^{-1}(1) \neq \tilde{\varphi}^{-1}\left(\frac{m_{+}\pi}{2N+1}\right)$, the function $$y\mapsto\frac{(k(y))^2 -1}{(k(y))^2 +1} \tan((2N+1)\tilde{\varphi}(y))$$ has exactly $2$ zeroes in both intervals $(y_{m_{-}}^+,y_{m_{-}}^- )$ and $(y_{m_{+}}^+,y_{m_{+}}^- )$. Hence,  the function $y\mapsto\frac{(k(y))^2 -1}{(k(y))^2 +1} \tan((2N+1)\tilde{\varphi}(y))$ has exactly $2N+4$ zeroes in the interval $[Y_{\mathrm{max}}^{2j},Y_{\mathrm{min}}^{2j}]$.

\noindent The rest of the proof does not change if $k^{-1}(-1) \in [Y_{\mathrm{max}}^{2j},y_{2N}^+ )$ or $k^{-1}(1) \in \left(y_1^-, Y_{\mathrm{min}}^{2j}\right] $.

\bigskip

\noindent \emph{Step 3.} We remark that 
$$\frac{2k(k^{-1}(\pm 1))}{1+(k(k^{-1}(\pm 1)))^2} = \pm 1$$
and 
$$\lim_{y\to (Y_{\mathrm{max}}^{2j})^{+}} \frac{2k(y)}{1+(k(y))^2} = \lim_{y\to (Y_{\mathrm{min}}^{2j})^{-}} \frac{2k(y)}{1+(k(y))^2} = 0. $$
Moreover, the function $y\mapsto  \frac{2k(y)}{1+(k(y))^2}$ is strictly decreasing on $[Y_{\mathrm{max}}^{2j},k^{-1}(-1)]$, strictly increasing on $[k^{-1}(-1), k^{-1}(1)]$ and strictly decreasing on $[k^{-1}(1),Y_{\mathrm{min}}^{2j} ]$. Hence, considering the monotonicity and the limits $\pm \infty$ of the function $$y\mapsto\frac{(k(y))^2 -1}{(k(y))^2 +1} \tan((2N+1)\tilde{\varphi}(y))$$ on each of the sub intervals $[Y_{\mathrm{max}}^{2j},Y_{\mathrm{min}}^{2j}]$ defined at Step 2, the function $$y\mapsto \frac{2k(y)}{1+(k(y))^2} + \frac{(k(y))^2 -1}{(k(y))^2 +1} \tan((2N+1)\tilde{\varphi}(y))$$ has the same number of zeroes as the function  $$y\mapsto\frac{(k(y))^2 -1}{(k(y))^2 +1} \tan((2N+1)\tilde{\varphi}(y))$$ on each of these sub intervals.  Hence, the function $$y\mapsto \frac{2k(y)}{1+(k(y))^2} + \frac{(k(y))^2 -1}{(k(y))^2 +1} \tan((2N+1)\tilde{\varphi}(y))$$ has exactly $2N+4$ zeroes in the interval $[Y_{\mathrm{max}}^{2j},Y_{\mathrm{min}}^{2j}]$.
\bigskip

\noindent \emph{Step 4.} The function $y\mapsto  \frac{1}{2^{2N+1}} ((a+\ii b)^{2N+1} + (a-\ii b)^{2N+1})(y+\cc)=2\cos((2N+1) \tilde{\varphi}(y+\cc))$ is continuous and does vanish on $[Y_{\mathrm{max}}^{2j},Y_{\mathrm{min}}^{2j}]$ only at the points $y_m^+$, $m\in \{0,\ldots,2N\}$. It does not vanish and is of constant sign on each interval $\left(y_m^+,y_m^- \right)$ for $m\in \{1,\ldots , 2N\}$. Moreover $b(Y_{\mathrm{max}}^{2j}+\cc)=b(Y_{\mathrm{min}}^{2j}+\cc) =0$ , $a(Y_{\mathrm{max}}^{2j}+\cc) <0$ and $a(Y_{\mathrm{min}}^{2j}+\cc) >0$. 

\noindent The function $y\mapsto (\alpha+\beta)(y+\cc)$ is strictly positive on the interval $(Y_{\mathrm{max}}^{2j},Y_{\mathrm{min}}^{2j})$. Moreover, $\beta(Y_{\mathrm{max}}^{2j}+\cc)=\beta(Y_{\mathrm{min}}^{2j}+\cc) =0$ and 
$$\lim_{y\to (Y_{\mathrm{max}}^{2j})^{+}} \alpha(y+\cc) = \lim_{y\to (Y_{\mathrm{min}}^{2j})^{-}} \alpha(y+\cc) = +\infty. $$ 


\noindent Hence, one already has that $\Phi$ has exactly $2N+2$ zeroes on the interval $(y_{2N}^+, y_1^- )$.

\noindent Since $y\mapsto \tan((2N+1)\tilde{\varphi}(y))$ is strictly negative on $(Y_{\mathrm{max}}^{2j},y_{2N}^+ )$ and vanishes at $Y_{\mathrm{max}}^{2j}$, one deduces from all the previous facts that either $\Phi(Y_{\mathrm{max}}^{2j}) >0$ or $\Phi(y)$ tends to $+\infty$ when $y$ tends to $(Y_{\mathrm{max}}^{2j})^{+}$. Since by definition of $\Phi$ it is a continuous function on $[-\cc,0]$, only the first case occurs. Hence, since $y\mapsto \cos((2N+1)\tilde{\varphi}(y))$ is strictly negative on $\left(Y_{\mathrm{max}}^{2j},y_{2N}^+ \right)$, $\Phi$ is strictly positive on this interval and does not vanish on it.

\noindent  Since $y\mapsto \tan((2N+1)\tilde{\varphi}(y))$ is strictly positive on $(y_{1}^-, Y_{\mathrm{min}}^{2j})$ and vanishes at $Y_{\mathrm{min}}^{2j}$, one deduces that either $\Phi(Y_{\mathrm{min}}^{2j}) >0$ or $\Phi(y)$ tends to $+\infty$ when $y$ tends to $(Y_{\mathrm{min}}^{2j})^{-}$ and since $\Phi$ is continuous, only the first case occurs. In particular, since $y\mapsto \cos((2N+1)\tilde{\varphi}(y))$ is strictly positive on the interval $(y_{1}^-, Y_{\mathrm{min}}^{2j})$,  $\Phi$ is strictly positive on this interval and thus does not vanish on it.

\noindent One concludes that $\Phi$ has exactly $2N+2$ zeroes in the interval  $[Y_{\mathrm{max}}^{2j},Y_{\mathrm{min}}^{2j}]$, which proves Theorem \ref{thm_spectre} in the case of an even spectral band. 

\bigskip

\noindent In the case of an odd spectral band, one has that $\tilde{\varphi}$ is a strictly increasing  homeomorphism from  $[Y_{\mathrm{max}}^{2j+1},Y_{\mathrm{min}}^{2j+1}]$ to $[0,\pi]$. Hence, Step 1 is the same as in the even case, replacing ``decreasing'' by ``increasing'', changing $-\infty$ into $+\infty$ and vice versa and taking care of the order of the numbers $y_m^{\pm}= \tilde{\varphi}^{-1}\left(\frac{(2m \pm 1)\pi}{2(2N+1)} \right)$ which is reversed. 

\noindent The function $\tilde{k}$ is a strictly decreasing function from $+\infty$ to $-\infty$ on $[Y_{\mathrm{max}}^{2j+1},Y_{\mathrm{min}}^{2j+1}]$, hence  Step 2 and Step 3 will not change except that the order of the numbers $\tilde{k}^{-1}(-1)$ and $\tilde{k}^{-1}(1)$ has to be changed.

\noindent Finally, Step 4 is similar, again inverting the ordering of the subintervals since $\tilde{\varphi}$ is now increasing. It finishes the proof of Theorem \ref{thm_spectre} in the case of an odd spectral band and thus in all cases.

\end{demo}

\section{Proof of Theorem \ref{thm_ids}}\label{sec_proof_ids}

The main ingredient needed to count the eigenvalues of $\HH_{2N+1}$ under a fixed real number is that we know from Section \ref{sec_proof_counting} that these eigenvalues are in the spectral bands of $\HH$, situated between two singularities of the function $\E \mapsto  \tan((2N+1)\varphi(\E))$. 

\noindent We also need more notations for the proof of Theorem \ref{thm_ids}. These notations will also be used in the proof of Lemma \ref{lem_variations_varphi}. Let $Ai$ and $Bi$ denote the usual Airy functions.
\vskip 2mm

\noindent \textbf{Notation.} We denote by $\{-a_j\}_{j\geq 1}$ the set of the zeroes of $Ai$ and by $\{-\tilde{a}_j\}_{j\geq 1}$ the set of the zeroes of $Ai'$ where the real numbers $-a_j$ and $-\tilde{a}_j$ are arranged in decreasing order. These sets are both subsets of $(-\infty,0]$. 

\noindent Let $j\geq 0$ and define the real numbers $\mathfrak{a}_{2j}= \tilde{a}_{j+1}$ and  $\mathfrak{a}_{2j+1}=a_{j+1}$.
\vskip 3mm

\begin{demo} (of Theorem \ref{thm_ids}).
We use the notations introduced in the proof of Theorem \ref{thm_spectre}. Let $\E \in [-\cc, 0]$.  

\noindent We start by counting the number of spectral bands of $\HH$ included in the interval $[-\cc, \E]$ and thus prove (\ref{eq_exp_pE}). Let $p\geq 0$. Using the results of \cite[Proposition 6.1]{BL}, the $p$-th spectral band of $\HH$ is centered at $-\cc+\mathfrak{a}_p$. Hence, the number $p(\E)$ of spectral bands of $\HH$ included in $[-\cc,\E]$ (of length $\cc+\E$) is the unique integer such that 
\begin{equation}\label{eq_ineg_cplusE}
\mathfrak{a}_{p(\E)}  \leq \cc + \E \leq \mathfrak{a}_{p(\E)+1}.
\end{equation}
With a similar argument as in the proof of \cite[Proposition 3.1]{BL}, we get, for every $j\geq 0$,
$$\frac{2}{3} (\tilde{a}_{j+1})^{\frac{3}{2}} \in \left[ \frac{\pi}{4}+j \pi - \frac{7}{36(\frac{\pi}{6}+j\pi)},  \frac{\pi}{4}+j \pi +\frac{7}{36(\frac{\pi}{6}+j\pi)} \right] $$
and
$$\frac{2}{3} (a_{j+1})^{\frac{3}{2}} \in \left[ \frac{3\pi}{4}+j \pi - \frac{5}{36(\frac{\pi}{6}+j\pi)},  \frac{3\pi}{4}+j \pi +\frac{5}{36(\frac{\pi}{6}+j\pi)} \right]. $$
Combined with (\ref{eq_ineg_cplusE}), it gives (\ref{eq_exp_pE}).

\bigskip 

\noindent We set, for every $m\in \{1,\ldots , 2N\}$, 
\begin{equation}\label{eq_def_em}
e_m^{\pm} =\varphi^{-1}\left( \frac{(2m\pm 1)\pi}{2(2N+1)} \right). 
\end{equation}

\bigskip
\noindent \emph{In an odd spectral band.} Let $j\geq 0$. Let $\E \in  [\E_{\mathrm{min}}^{2j+1},\E_{\mathrm{max}}^{2j+1} ]$. From Proposition \ref{prop_variations_arg}, $\varphi$  defined at (\ref{def_argument}) having the same variations on $[\E_{\mathrm{min}}^{p},\E_{\mathrm{max}}^{p} ]$ as $\tilde{\varphi}$ on $[Y_{\mathrm{max}}^{p},Y_{\mathrm{min}}^{p} ]$, it is strictly increasing on $  [\E_{\mathrm{min}}^{2j+1},\E_{\mathrm{max}}^{2j+1} ]$. 

\noindent Let $m_{\E}\in \{1,\ldots, 2N \}$ be the unique integer such $\E \in [ e_{m_{\E}}^{-},e_{m_{\E}}^{+}  ]$ and let $m_{\E}=0$ if $\E \in  [ \E_{\mathrm{min}}^{2j+1} ,  e_1^- )$ and $m_{\E}=2N$ if $\E \in  (  e_{2N}^+,  \E_{\mathrm{max}}^{2j+1} ]$.

\noindent From the proof of Theorem \ref{thm_spectre}, we already know that in each interval $(e_m^-,e_m^+ )$ for $m\in \{1,\ldots , 2N\}$, there is exactly one eigenvalue of $\HH$, except for $m=m_{\pm}$ for which there are $2$ eigenvalues of $\HH$. We also know that there is no eigenvalues of $\HH$ in $[ \E_{\mathrm{min}}^{2j+1} , e_1^- )$ and in $(  e_{2N}^+,  \E_{\mathrm{min}}^{2j+1}]$ which explains the definition of $m_{\E}$. Hence, there are exactly $m_{\E}+n_{\E}$ eigenvalues of $\HH$  in the interval $[\E_{\mathrm{min}}^{2j+1}, \E]$ with $n_{\E}\in \{ 0, 1, 2 \}$. 

\noindent But, since $\varphi$ is strictly increasing, 
$$\frac{(2m_{\E}-1)\pi}{2(2N+1)}  \leq \varphi(\E) \leq \frac{(2m_{\E}+1)\pi}{2(2N+1)} $$
and
\begin{equation}\label{eq_ineg_thm2_mE_1}
\frac{2N+1}{\pi} \varphi(\E) - \frac12 \leq m_{\E} \leq  \frac{2N+1}{\pi} \varphi(\E) + \frac12.
\end{equation}
Hence, 
\begin{equation}\label{eq_demo_thm2_conv_odd}
\lim_{N\to +\infty} \frac{1}{2(2N+1)} (m_{\E}+ n_{\E}) = \frac{1}{2\pi} \varphi(\E).
\end{equation}

\bigskip 

\noindent \emph{In an even spectral band.} The argument is similar to the odd case, except that $\varphi$ is strictly decreasing as a consequence of Proposition \ref{prop_variations_arg}. Let $j\geq 0$ and let $\E \in   [\E_{\mathrm{min}}^{2j},\E_{\mathrm{max}}^{2j} ]$. We define $m_{\E}$ as in the odd case except that we exchange the bounds of each subinterval of $ [\E_{\mathrm{min}}^{2j},\E_{\mathrm{max}}^{2j} ]$ and we set $m_{\E}=2N$ if $\E \in  [ \E_{\mathrm{min}}^{2j} ,  e_{2N}^+ )$ and $m_{\E}=0$ if $\E \in  [  e_1^-,  \E_{\mathrm{max}}^{2j} ]$. 

\noindent Since $\varphi$ is decreasing, as in the odd case, the number of eigenvalues of $\HH$ in $[\E,  \E_{\mathrm{max}}^{2j}]$ is equal to $m_{\E}+n_{\E}$ with $n_{\E}\in \{ 0, 1, 2 \}$. Hence, using the result of Theorem \ref{thm_spectre}, there are exactly $2N+2 - (m_{\E}+n_{\E})$ eigenvalues of $\HH$ in the interval $[\E_{\mathrm{min}}^{2j}, \E]$. 

\noindent We still have an inequality similar to (\ref{eq_ineg_thm2_mE_1}) except that we exchange the upper and lower bounds, thus
\begin{equation}\label{eq_demo_thm2_conv_even}
\lim_{N\to +\infty} \frac{1}{2(2N+1)} (2N+2-(m_{\E}+ n_{\E})) = \frac{1}{2}-\frac{1}{2\pi} \varphi(\E).
\end{equation}
\bigskip 

\noindent \emph{Conclusion.} Since $\varphi$ is an homeomorphism from each spectral band to $[0,\pi]$, from (\ref{eq_demo_thm2_conv_even}) and (\ref{eq_demo_thm2_conv_odd}), we get that if $\E$ is in a spectral gap of $\HH$, the integrated density of states of $\HH$, $\mathbf{I}(\E)$, is equal to $\frac{1}{2} p(\E)$, where $p(\E)$ is the number of spectral bands included in $[-\cc, \E]$.  Theorem \ref{thm_ids} follows as this remark,  (\ref{eq_demo_thm2_conv_even}) and (\ref{eq_demo_thm2_conv_odd}) implies (\ref{eq_exp_ids}) and (\ref{eq_exp_pE}) has already been proven.
\end{demo}

\appendix

\section{Proof of Lemma \ref{lem_variations_varphi}}\label{sec_appendix_lemma_h}

To prove Lemma \ref{lem_variations_varphi}, we use well-known properties of the classical Airy functions $Ai$ and $Bi$ and their approximations combined with results from \cite{BL}. The idea is the following : the $p$-th spectral band of $\HH$ is centered at $-\cc+\mathfrak{a}_p$, hence we prove that the value of $h(-\mathfrak{a}_p)$ is strictly positive when $p$ is even and strictly negative when $p$ is odd. Since the length of each spectral band is exponentially small, it implies, using Taylor formula, that the sign of $h$ does not change on each spectral band. We focus on the case $p$ even, the odd case is completely similar.
\bigskip

\noindent To start, note that one has the expression of $u$ and $v$ in terms of $Ai$ and $Bi$:
$$\forall x \in \R,\ u(x)=\pi(Bi'(0)Ai(x)-Ai'(0)Bi(x))$$ 
and
$$\forall x \in \R,\   v(x)=\pi(Ai(0)Bi(x)-Bi(0)Ai(x)).$$
Also recall that the Wronskian of $Ai$ and $Bi$ is equal to $\frac{1}{\pi}$. Let $j\geq 0$, and $\cc \geq c_{2j}$. Since $Ai'(-\tilde{a}_{j+1})=0$,
\begin{align}
\U(-\tilde{a}_{j+1}+\cc) & = \pi Bi'(-\tilde{a}_{j+1}) Ai(-\tilde{a}_{j+1}+\cc) ; \label{eq_value_U_atc} \\
\U'(-\tilde{a}_{j+1}+\cc) & = \pi Bi'(-\tilde{a}_{j+1}) Ai'(-\tilde{a}_{j+1}+\cc) ; \label{eq_value_Up_atc} \\
\V(-\tilde{a}_{j+1}+\cc) & = \pi (Ai(-\tilde{a}_{j+1})Bi(-\tilde{a}_{j+1})-Bi(-\tilde{a}_{j+1}) Ai(-\tilde{a}_{j+1}+\cc)) ; \label{eq_value_V_atc} \\
\V'(-\tilde{a}_{j+1}+\cc) & = \pi (Ai(-\tilde{a}_{j+1})Bi'(-\tilde{a}_{j+1}) - Bi(-\tilde{a}_{j+1})Ai'(-\tilde{a}_{j+1}+\cc)). \label{eq_value_Vp_atc} 
\end{align}
Hence,
\begin{align}
h(-\tilde{a}_{j+1})& = \pi^2 \left[ (\tilde{a}_{j+1}(Bi(-\tilde{a}_{j+1}))^2 -(Bi'(-\tilde{a}_{j+1}))^2  ) (AiAi')(-\tilde{a}_{j+1}+\cc) \right. \label{eq_expr_h_Airy1} \\ 
& + \tilde{a}_{j+1}(Ai(-\tilde{a}_{j+1}))^2 (BiBi')(-\tilde{a}_{j+1}+\cc)\label{eq_expr_h_Airy2} \\
& -\frac{1}{\pi} (AiBi)(-\tilde{a}_{j+1})  \label{eq_expr_h_Airy3} \\
& + (AiBi')(-\tilde{a}_{j+1})(Ai'Bi')(-\tilde{a}_{j+1}+\cc) \label{eq_expr_h_Airy4} \\
& -(BiBi')(-\tilde{a}_{j+1}) (Ai'(-\tilde{a}_{j+1}+\cc))^2   \label{eq_expr_h_Airy5} \\
& + (-\tilde{a}_{j+1}+\cc) (AiBi')(-\tilde{a}_{j+1})(AiBi)(-\tilde{a}_{j+1}+\cc) \label{eq_expr_h_Airy6}  \\
& \left.-(-\tilde{a}_{j+1}+\cc) (BiBi')(-\tilde{a}_{j+1}) (Ai(-\tilde{a}_{j+1}+\cc))^2 \right]. \label{eq_expr_h_Airy7}
\end{align}
One already sees that in the expression of $h(-\tilde{a}_{j+1})$, the dominant term when $\cc$ is large is (\ref{eq_expr_h_Airy2}), hence positive. But our result is stated not for $\cc$ arbitrarily large, but for every $\cc \geq c_{2j}$. Since it is increasing in $\cc$, we may assume in the sequel that $\cc = c_{2j}$ and the result follows for any $\cc \geq c_{2j}$.

\noindent First, for $j=0$, one has $-\tilde{a}_1 \simeq -1.088$, $c_0\simeq 1.515$ and $h(-\tilde{a}_1) \simeq 1.428 >0$. In the sequel, we assume $j\geq 1$.

\noindent Using \cite[10.4.95]{AS}, 
\begin{equation}\label{eq_demo_lemma1_approx01}
\left| -\tilde{a}_{j+1} - \left( -\left( \frac{3\pi}{8}(4j+3) \right)^{\frac23}  \right) \right| \leq \frac{5}{48}  \left( \frac{3\pi}{8}(4j+3) \right)^{-\frac43}.
\end{equation}
Using \cite[10.4.97]{AS},
 \begin{equation}\label{eq_demo_lemma1_approx02}
\left| Ai(-\tilde{a}_{j+1}) -  \frac{(-1)^j}{\sqrt{\pi}} \left( \frac{3\pi}{8}(4j+3) \right)^{\frac16}  \right| \leq \frac{5}{48} \frac{1}{\sqrt{\pi}}  \left( \frac{3\pi}{8}(4j+3) \right)^{-\frac{11}{6}}.
\end{equation}
By \cite[10.4.64]{AS} and (\ref{eq_demo_lemma1_approx01}),
 \begin{equation}\label{eq_demo_lemma1_approx03}
\left| Bi(-\tilde{a}_{j+1}) -  \frac{(-1)^{j+1}}{\sqrt{\pi}} \left( \frac{3\pi}{8}(4j+3) \right)^{-\frac16}  \right| \leq \frac{385}{4608} \frac{1}{\sqrt{\pi}} \left( \frac{3\pi}{8}(4j+3)   \right)^{-\frac{13}{6}}.
\end{equation}
Using \cite[10.4.67]{AS},
 \begin{equation}\label{eq_demo_lemma1_approx04}
\left| Bi'(-\tilde{a}_{j+1}) - \frac{3}{2}  \frac{(-1)^{j}}{\sqrt{\pi}} \left( \frac{3\pi}{8}(4j+3) \right)^{-\frac54}  \right| \leq \frac{7315}{663552} \frac{1}{\sqrt{\pi}} \left( \frac{3\pi}{8}(4j+3)   \right)^{-\frac{17}{6}}.
\end{equation}
Combining (\ref{eq_demo_lemma1_approx01}) and \cite[Proposition 3.1]{BL}, one also has
\begin{align}
\left| (-\tilde{a}_{j+1}+c_{2j}) - \frac29 \left( \frac{3\pi}{2} \right)^{\frac23} j^{-\frac13} \right| & \leq   \frac{5}{48}  \left( \frac{3\pi}{8}(4j+3) \right)^{-\frac43} + \left( \frac{7}{8\pi \left( j+\frac13 \right)} \right)^{\frac23} \nonumber \\
& \leq  2 \left( \frac{3\pi}{8}(4j+3) \right)^{-\frac23}.\label{eq_demo_lemma1_approx05}
\end{align}
Note that
$$ 2 \left( \frac{3\pi}{8}(4j+3) \right)^{-\frac23} \leq \frac29 \left( \frac{3\pi}{2} \right)^{\frac23} j^{-\frac13} + 2 \left( \frac{3\pi}{8}(4j+3) \right)^{-\frac23} \leq \frac49 \left( \frac{3\pi}{2} \right)^{\frac23} j^{-\frac13}.$$
Hence 
\begin{equation}\label{eq_demo_lemma1_approx06}
\frac{1}{\pi} \frac{8}{27} j^{\frac12} \leq \frac32 \left(  \frac29 \left( \frac{3\pi}{2} \right)^{\frac23} j^{-\frac13} + 2 \left( \frac{3\pi}{8}(4j+3) \right)^{-\frac23}\right)^{-\frac32} \leq \frac{9\pi}{32\sqrt{2}} (4j+3),
\end{equation}

\begin{equation}\label{eq_demo_lemma1_approx07}
\frac{1}{2\sqrt{2}} (4j+3)^{-1} \leq \frac43 \left(  \frac29 \left( \frac{3\pi}{2} \right)^{\frac23} j^{-\frac13} + 2 \left( \frac{3\pi}{8}(4j+3) \right)^{-\frac23}\right)^{\frac32} \leq \frac{16}{27} \pi j^{-\frac12}, 
\end{equation}

\begin{equation}\label{eq_demo_lemma1_approx08}
\left(  \frac29 \left( \frac{3\pi}{2} \right)^{\frac23} j^{-\frac13} + 2 \left( \frac{3\pi}{8}(4j+3) \right)^{-\frac23}\right)^{\frac12} \leq \frac23 \left( \frac{3\pi}{2} \right)^{\frac13} j^{-\frac16} 
\end{equation}
and
\begin{equation}\label{eq_demo_lemma1_approx09}
\left(  \frac29 \left( \frac{3\pi}{2} \right)^{\frac23} j^{-\frac13} + 2 \left( \frac{3\pi}{8}(4j+3) \right)^{-\frac23}\right)^{-\frac12} \leq \frac{1}{\sqrt{2}} \left( \frac{3\pi}{8}(4j+3) \right)^{\frac13}.
\end{equation}
Using \cite[10.4.59 and 10.4.61]{AS}, (\ref{eq_demo_lemma1_approx06}) and (\ref{eq_demo_lemma1_approx07}),
\begin{equation}\label{eq_demo_lemma1_approx10}
\left| (Ai Ai')(-\tilde{a}_{j+1}+c_{2j}) - \left( -\frac{1}{4\pi} \ee^{-2\pi \left( \frac29 \right)^{\frac32} j^{-\frac12} } \right)  \right| \leq \frac{5}{1024\sqrt{2}} (4j+3) \ee^{-\frac{1}{2\sqrt{2}} (4j+3)^{-1}} .
\end{equation}
Using \cite[10.4.63 and 10.4.66]{AS}, (\ref{eq_demo_lemma1_approx06}) and (\ref{eq_demo_lemma1_approx07}),
\begin{equation}\label{eq_demo_lemma1_approx11}
\left| (Bi Bi')(-\tilde{a}_{j+1}+c_{2j}) - \frac{1}{\pi} \ee^{2\pi \left( \frac29 \right)^{\frac32} j^{-\frac12} } \right| \leq \frac{5}{1024\sqrt{2}} (4j+3) \ee^{ \frac{16}{27} \pi j^{-\frac12} } .
\end{equation}
Using \cite[10.4.61 and 10.4.66]{AS}, (\ref{eq_demo_lemma1_approx06}) and (\ref{eq_demo_lemma1_approx08}),
\begin{equation}\label{eq_demo_lemma1_approx12}
\left| (Ai' Bi')(-\tilde{a}_{j+1}+c_{2j}) - \left( -\frac{1}{2\pi} \left( \frac29 \right)^{\frac12} \left( \frac{3\pi}{2} \right)^{\frac13} j^{-\frac16}  \right) \right| \leq \frac{7}{768\sqrt{2}}\left( \frac{3\pi}{2}\right)^{\frac13} j^{-\frac16} (4j+3).
\end{equation}
Using \cite[10.4.61]{AS}, (\ref{eq_demo_lemma1_approx06}), (\ref{eq_demo_lemma1_approx07}) and  (\ref{eq_demo_lemma1_approx08}),
\begin{equation}\label{eq_demo_lemma1_approx13}
\left| (Ai'(-\tilde{a}_{j+1}+c_{2j}))^2 - \frac{1}{4\pi} \left( \frac29 \right)^{\frac12} \left( \frac{3\pi}{2} \right)^{\frac13} j^{-\frac16} \ee^{-2\pi \left( \frac29 \right)^{\frac32} j^{-\frac12} }   \right| \leq \frac{7}{1536}  \left( \frac{3\pi}{2} \right)^{\frac13}  j^{-\frac16} (4j+3) \ee^{-\frac{1}{2\sqrt{2}} (4j+3)^{-1}}.
\end{equation}
Using \cite[10.4.59 and 10.4.63]{AS}, (\ref{eq_demo_lemma1_approx06}) and (\ref{eq_demo_lemma1_approx09}),
\begin{equation}\label{eq_demo_lemma1_approx14}
\left| (Ai Bi)(-\tilde{a}_{j+1}+c_{2j}) - \frac{1}{2\pi} \left( \frac29 \right)^{-\frac12} \left( \frac{3\pi}{2} \right)^{-\frac13} j^{\frac16}  \right| \leq \frac{5}{192}\left( \frac{3\pi}{8}(4j+3) \right)^{\frac43}.
\end{equation}
Finally, using \cite[10.4.59]{AS}, (\ref{eq_demo_lemma1_approx06}), (\ref{eq_demo_lemma1_approx07}) and  (\ref{eq_demo_lemma1_approx09}),
\begin{equation}\label{eq_demo_lemma1_approx15}
\left| (Ai(-\tilde{a}_{j+1}+c_{2j}))^2 - \frac{1}{4\pi} \left( \frac29 \right)^{-\frac12} \left( \frac{3\pi}{2} \right)^{-\frac13} j^{\frac16} \ee^{-2\pi \left( \frac29 \right)^{\frac32} j^{-\frac12} }   \right| \leq \frac{5}{768} \frac{1}{\pi} \left( \frac{3\pi}{8}(4j+3) \right)^{\frac43} \ee^{-\frac{1}{2\sqrt{2}} (4j+3)^{-1}}.
\end{equation}
With (\ref{eq_demo_lemma1_approx01}), (\ref{eq_demo_lemma1_approx03}), (\ref{eq_demo_lemma1_approx04}) and (\ref{eq_demo_lemma1_approx10}) we estimate (\ref{eq_expr_h_Airy1}). With (\ref{eq_demo_lemma1_approx01}), (\ref{eq_demo_lemma1_approx02}) and (\ref{eq_demo_lemma1_approx11}) we estimate (\ref{eq_expr_h_Airy2}). With (\ref{eq_demo_lemma1_approx02}) and (\ref{eq_demo_lemma1_approx03}) we estimate (\ref{eq_expr_h_Airy3}). With (\ref{eq_demo_lemma1_approx02}), (\ref{eq_demo_lemma1_approx04}) and (\ref{eq_demo_lemma1_approx12}) we estimate (\ref{eq_expr_h_Airy4}).  With (\ref{eq_demo_lemma1_approx03}), (\ref{eq_demo_lemma1_approx04}) and (\ref{eq_demo_lemma1_approx13}) we estimate (\ref{eq_expr_h_Airy5}).  With (\ref{eq_demo_lemma1_approx05}), (\ref{eq_demo_lemma1_approx02}), (\ref{eq_demo_lemma1_approx04}) and (\ref{eq_demo_lemma1_approx14}) we estimate (\ref{eq_expr_h_Airy6}). At last, with (\ref{eq_demo_lemma1_approx05}), (\ref{eq_demo_lemma1_approx03}), (\ref{eq_demo_lemma1_approx04}) and (\ref{eq_demo_lemma1_approx15}) we estimate (\ref{eq_expr_h_Airy7}). After tedious computations, we obtain, for every $j\geq 1$ and for $\cc = c_{2j}$,
\begin{align}
& \left| h(-\tilde{a}_{j+1})   - \left( 1+ \frac{3\pi}{8}(4j+3) \ee^{2\pi \left( \frac29 \right)^{\frac32} j^{-\frac12}} - \frac{3\pi}{32}(4j+3) \left( 1-\frac94 \left( \frac{3\pi}{8}(4j+3) \right)^{-\frac72}\right) \ee^{-2\pi \left( \frac29 \right)^{\frac32} j^{-\frac12}} \right. \right. \nonumber \\
&\left. \left. \quad  + \frac58 \left( \frac29 \right)^{\frac12} \left( \frac{3\pi}{2} \right)^{\frac13} j^{-\frac16}\left( \frac{3\pi}{8}(4j+3) \right)^{-\frac{35}{24}}  \ee^{-2\pi \left( \frac29 \right)^{\frac32} j^{-\frac12}} \right)\right| \nonumber \\
& \leq \pi \left( \left( \frac{3\pi}{8}(4j+3) \right)^{-\frac{14}{3}} \frac{8}{3\pi}  \ee^{-\frac{1}{2\sqrt{2}} (4j+3)^{-1}} \left( \frac{5}{48} \left( \frac{385}{4608} \right)^2 + \left( \frac{7315}{663552} \right)^2 \right)\frac{5}{1024\sqrt{2}} \right. \nonumber \\
& + \left( \frac{3\pi}{8}(4j+3) \right)^{-4} \frac{8}{3\pi}  \ee^{ \frac{16}{27} \pi j^{-\frac12} } \left(\frac{5}{48}\right)^3 \frac{5}{1024\sqrt{2}} + \frac{1}{\pi} \left( \frac{3\pi}{8}(4j+3) \right)^{-4} \frac{5}{48} \frac{385}{4608} \nonumber \\
& + \left( \frac{3\pi}{8}(4j+3) \right)^{-\frac{11}{3}}    j^{-\frac16} \frac{5}{48}\frac{7315}{663552} \frac{7}{768\sqrt{2}} \frac{8}{3\pi} \left( \frac{3\pi}{2} \right)^{\frac13} \nonumber \\
& + \left( \frac{3\pi}{8}(4j+3) \right)^{-4} j^{-\frac16} \ee^{-\frac{1}{2\sqrt{2}} (4j+3)^{-1}}  \frac{385}{4608}\frac{7315}{663552} \frac{7}{1536} \frac{8}{3\pi} \left( \frac{3\pi}{2} \right)^{\frac13} \nonumber \\
& \left. +  \left( \frac{3\pi}{8}(4j+3) \right)^{-4}  \frac{5}{24} \frac{7315}{663552} \frac{5}{192} +      \left( \frac{3\pi}{8}(4j+3) \right)^{-\frac{13}{3}}  \ee^{-\frac{1}{2\sqrt{2}} (4j+3)^{-1}} \frac{2}{\pi}  \frac{385}{4608}\frac{7315}{663552} \frac{5}{768} \right). \label{eq_demo_lemma1_approx_finale1}
\end{align}
In the upper bound of (\ref{eq_demo_lemma1_approx_finale1}), the dominant term is, for every $j\geq 1$, $\frac{1}{\pi} \left( \frac{3\pi}{8}(4j+3) \right)^{-4} \frac{5}{48} \frac{385}{4608}$. Since $\frac{5}{48} \frac{385}{4608} \leq 9\times 10^{-3}$, one finally obtains the less precise but sufficient estimate :
\begin{align}
& \left| h(-\tilde{a}_{j+1})   - \left( 1+ \frac{3\pi}{8}(4j+3) \ee^{2\pi \left( \frac29 \right)^{\frac32} j^{-\frac12}} - \frac{3\pi}{32}(4j+3) \left( 1-\frac94 \left( \frac{3\pi}{8}(4j+3) \right)^{-\frac72}\right) \ee^{-2\pi \left( \frac29 \right)^{\frac32} j^{-\frac12}} \right. \right. \nonumber \\
&\left. \left. \quad  + \frac58 \left( \frac29 \right)^{\frac12} \left( \frac{3\pi}{2} \right)^{\frac13} j^{-\frac16}\left( \frac{3\pi}{8}(4j+3) \right)^{-\frac{35}{24}}  \ee^{-2\pi \left( \frac29 \right)^{\frac32} j^{-\frac12}} \right)\right| \nonumber \\
& \leq 7\times 9 \times 10^{-3}  \left( \frac{3\pi}{8}(4j+3) \right)^{-4}.  \label{eq_demo_lemma1_approx_finale2}
\end{align}
The approximation of $h(-\tilde{a}_{j+1})$ in (\ref{eq_demo_lemma1_approx_finale2}) is increasing in $j$ and is approximately equal to $15.87$ for $j=1$, hence for every $j\geq 1$,
\begin{equation}\label{eq_demo_lemma1_approx_finale3}
 h(-\tilde{a}_{j+1}) \geq 15.87 - 6.3 \times 10^{-2} \times   \left( \frac{3\pi}{8}(4j+3) \right)^{-4} >0. 
\end{equation}
We prove that $h$ remains strictly positive on the spectral band $[Y_{\mathrm{max}}^{2j},Y_{\mathrm{min}}^{2j}]$. Recall that the spectral bands of $\HH$ are exponentially small. More precisely, \cite[Theorem 2.5]{BL} implies that, for every $\cc \geq c_{2j}$,
\begin{equation}\label{eq_inclusion_bande}
[Y_{\mathrm{max}}^{2j},Y_{\mathrm{min}}^{2j}] \subset  \left[ -\tilde{a}_{j+1}-\Lambda_{2j,\cc},  -\tilde{a}_{j+1}+\Lambda_{2j,\cc} \right].
\end{equation}
where 
\begin{equation}\label{eq_def_Lc2j}
\Lambda_{2j,\cc}:=  \left( \frac{(Bi'(-\tilde{a}_{j+1}))^2}{2\pi\tilde{a}_{j+1}} + \frac{K_{2j}}{(\cc-\tilde{a}_{j+1})^{\frac32}} \right) \ee^{-\frac43 (\cc-\tilde{a}_{j+1})^{\frac32} }
\end{equation}
and where we used the relationships $u'(-\tilde{a}_{j+1})=-Ai'(0)Bi'(-\tilde{a}_{j+1})$ and $Ai(0)Ai'(0)=-\frac{1}{2\pi \sqrt{3}}$. Note that the notations in  \cite{BL} and in the present paper correspond through $\cc=\mathsf{h}^{-\frac23}$. 

\noindent The constant $K_{2j}$ is defined in the proof of \cite[Proposition 6.1]{BL}, but it can be much improved for our purpose. Indeed, using the notations given in this proof, one can actually set, for every $\tau>0$, 
$$B_{2j,\tau}:=\sqrt{3} \ee^{\frac23 \tau^{-\frac12} (c_{2j}-\tilde{a}_{j+1})^2 }(1+2.83\sqrt{3} \tau^{-\frac32} )$$
instead of
$$B_{2j,\tau}:=\sqrt{3} \ee^{\frac23 \tau^{-\frac12} \tilde{a}_{j+1}^2 }(1+2.83\sqrt{3} \tau^{-\frac32} ).$$
Indeed, similar to \cite[(4.3) or (4.22)]{BL}, one actually set $X=-X_{\mathrm{max}}^{2j}-\mathsf{h}^{-\frac23}+\tilde{a}_{j+1} \in [-c_{2j}+\tilde{a}_{j+1},0)$ which gives the good estimate corresponding to \cite[(4.11)]{BL}. 

\noindent Following the proof of \cite[Proposition 6.1]{BL} and using (\ref{eq_demo_lemma1_approx05}), we get that, for $\tau=c_{2j}-\tilde{a}_{j+1}$,
\begin{equation}\label{eq_upper_bound_K2j}
\forall j\geq 1,\ |K_{2j}|\leq 2\times 10^6 \times j^{-\frac{11}{6}} \ee^{4\pi \left( \frac29 \right)^{\frac32} j^{-\frac12}}.  
\end{equation}
Thus, using (\ref{eq_demo_lemma1_approx01}) and (\ref{eq_demo_lemma1_approx04}), for every $j\geq 1$,
\begin{align}
\left|  \frac{(Bi'(-\tilde{a}_{j+1}))^2}{2\pi\tilde{a}_{j+1}} + \frac{K_{2j}}{(c_{2j}-\tilde{a}_{j+1})^{\frac32}} \right| & \leq \frac{9}{2} \frac{1}{\pi^2} \left( \frac{3\pi}{8}(4j+3) \right)^{-\frac{19}{6}} + 2\times 10^6 \times j^{-\frac{11}{6}} \ee^{4\pi \left( \frac29 \right)^{\frac32} j^{-\frac12}}. \nonumber \\
& \leq 3\times 10^6 \times j^{-\frac{11}{6}} \ee^{4\pi \left( \frac29 \right)^{\frac32} j^{-\frac12}}. \label{eq_upper_bound_lemma1_bands}
\end{align}
Hence, using (\ref{eq_demo_lemma1_approx08}), for every $j\geq 1$,
\begin{equation}\label{eq_upper_bound_lemma1_bands2}
\left| \Lambda_{2j,c_{2j}} \right| \leq 3\times 10^6 \times j^{-\frac{11}{6}} \ee^{4\pi \left( \frac29 \right)^{\frac32}j^{-\frac12} -\frac{1}{2\sqrt{2}} (4j+3)^{-1}} .
\end{equation}
Let $t\in[-1,1]$. Since $h$ is $C^1$, by the Mean Value Theorem, for every $j\geq 1$ and every $\cc \geq c_{2j}$,
\begin{equation}\label{eq_TAF_h_band}
\left| h(-\tilde{a}_{j+1} + t \Lambda_{2j,\cc}) - h(-\tilde{a}_{j+1})  \right| \leq \left( \sup_{y\in [Y_{\mathrm{max}}^{2j},Y_{\mathrm{min}}^{2j}]} \left| \frac{\dd h}{\dd y}(y) \right| \right) \times |\Lambda_{2j,\cc}|
\end{equation}
But, using (\ref{eq_derivee_U}), (\ref{eq_derivee_Up}), (\ref{eq_derivee_V}) and (\ref{eq_derivee_Vp}), for every $y$,
\begin{align}
\frac{\dd h}{\dd y}(y) & = 2(2y+\cc)(\U \V' + \V \U')(y+\cc) - ((\U'(y+\cc))^2 + (y+\cc)\U(y+\cc))\nonumber \\
 & - 2y ((\V'(y+\cc))^2 + (y+\cc)\V(y+\cc)) + (\V (\U - \V'))(y+\cc).\label{eq_deriv_h}
\end{align}
Using \cite[10.4.3, 10.4.63,10.4.66]{AS}, if $\eta>0$ is any real number,
\begin{equation}\label{eq_ineg_BiBip}
\forall x\geq \left( \frac{Ai'(0)}{Ai(0)} \right)^2 \eta^{-2},\ 0\leq Bi'(x) \leq \eta x^{\frac12} Bi(x).  
\end{equation}
Since for any $x\geq 0$, $Ai(x)\leq Bi'(x)$ and $|Ai'(x)|\leq Bi'(x)$, using the expressions of $u$ and $v$ in terms of Airy functions leads to, for any real number $\eta>0$ and any $y\in [-\cc,0]$,
\begin{align}
y+\cc \geq    \left( \frac{Ai'(0)}{Ai(0)} \right)^2 \eta^{-2} & \Rightarrow \ |\U'(y+\cc)| \leq \eta (y+\cc)^{\frac12} |\U(y+\cc)| \label{ineg_UpU} \\
y+\cc \geq \left( \frac{Ai'(0)}{Ai(0)} \right)^2 \eta^{-2} & \Rightarrow \ |\V'(y+\cc)| \leq \eta (y+\cc)^{\frac12} |\V(y+\cc)|. \label{ineg_VpV}
\end{align}
Moreover, for every $y\in [-\cc,0]$,
\begin{equation}\label{ineg_UUp}
|\U(y+\cc)| \leq 2 |\U'(y+\cc)| 
\end{equation} 
and
\begin{equation}\label{ineg_VVp}
|\V(y+\cc)| \leq 2 |\V'(y+\cc)|.
\end{equation} 
If $y\in [Y_{\mathrm{max}}^{2j},Y_{\mathrm{min}}^{2j}]$, then $y+\cc \geq Y_{\mathrm{max}}^{2j}+\cc$. Hence, using (\ref{ineg_UpU}) and (\ref{ineg_VpV}) with $\eta=\left( \frac{Ai'(0)}{Ai(0)}\right)^{\frac12} (Y_{\mathrm{max}}^{2j}+\cc)^{-\frac12}$ and using (\ref{ineg_UUp}) and (\ref{ineg_VVp}), 
\begin{equation}\label{eq_ineg_h_hp}
  \sup_{y\in [Y_{\mathrm{max}}^{2j},Y_{\mathrm{min}}^{2j}]} \left| \frac{\dd h}{\dd y}(y) \right|  \leq 6 \left( \frac{Ai'(0)}{Ai(0)}\right)^{\frac12} (Y_{\mathrm{max}}^{2j}+\cc)^{-\frac12} (Y_{\mathrm{min}}^{2j}+\cc)^{\frac32} \sup_{y\in [Y_{\mathrm{max}}^{2j},Y_{\mathrm{min}}^{2j}]} \left| h(y) \right| 
\end{equation}
Combining (\ref{eq_demo_lemma1_approx_finale2}), (\ref{eq_demo_lemma1_approx_finale3}), (\ref{eq_upper_bound_lemma1_bands2}), (\ref{eq_TAF_h_band}) and (\ref{eq_ineg_h_hp}), one gets that (for $\cc=c_{2j}$), for every $j\geq 1$ and every $t\in [-1,1]$, $ h(-\tilde{a}_{j+1} + t \Lambda_{2j,c_{2j}}) >0$. Therefore, for every $\cc \geq c_{2j}$, every $j\geq 1$ and every $t\in [-1,1]$, $ h(-\tilde{a}_{j+1} + t \Lambda_{2j,\cc}) >0$. By (\ref{eq_inclusion_bande}), it leads to $h(y)>0$ for every $y\in [Y_{\mathrm{max}}^{2j},Y_{\mathrm{min}}^{2j}]$. This proves Lemma \ref{lem_variations_varphi}. 
\bigskip

\noindent \textbf{Remark.} With (\ref{eq_inclusion_bande}), (\ref{eq_def_Lc2j}) and (\ref{eq_TAF_h_band}) it is easy to show that for $\cc$ arbitrary large, the result of Lemma \ref{lem_variations_varphi} is true. What is difficult here is to be able to take $\cc$ only larger than $c_{2j}$. 







\section{The case of an even number of wells}\label{sec_appendix_M_even}

All the results remain true in the case of an even number of potential wells. Slight changes have to be made in order to carry on the proofs done in the odd case. First, we have to define the potential corresponding to an even number $2N$ of wells in a way that the potential function is an even function.
\begin{equation}
 V_{2N} : x\mapsto \sum_{k=-N+1}^N V(x-2kL_0 + L_0)
\end{equation}
where $V$ is defined in (\ref{eq_def_V}). After rescaling as in the odd case, we consider
\[ \mathbf{V}_{2N} : z\mapsto \sum_{k=-N+1}^N \mathbf{V}(z-2k + 1)\] 
Then, instead of defining the functions $\U$ and $\V$ as in (\ref{def_U}) and (\ref{def_V}), we set : 
\[
 \U(x)=v'(- \E)u(x)-u'(- \E)v(x)
\]
and
\[
\V(x)= u( - \E)v(x)-v(- \E)u(x).
\]

The parity of the number of wells played a role in the proof of the absence of eigenvalues in the spectral gaps of $\HH$. Following the proof of Section \ref{sec_no_evs_gaps}, there is no change to do in the first case. In the second case, corresponding to the spectral gaps $(Y_{\mathrm{min}}^{4j+3},Y_{\mathrm{max}}^{4j+2})$ or $(Y_{\mathrm{min}}^{4j+1},Y_{\mathrm{max}}^{4j})$, one has still $a+b<0$ and $a-b<0$, but now $(a+b)^{2N}>0$ and $(a-b)^{2N}>0$. Thus, the first term in (\ref{eq_expr_Phi_gaps}) is strictly positive. For the second term, since $a-b<a+b<0$ we have $(a-b)^{2N}>(a+b)^{2N}$ and the second term in (\ref{eq_expr_Phi_gaps}) is now strictly negative. In particular, it does not vanish on these spectral gaps and thus $H_{2N}$ has no eigenvalues in the spectral gaps of $H$.

In the other proofs, the parity of the number of wells does not play any role. Our choice of focusing on an odd number of wells was guided by the fact that it is naturally involved in the definition of the integrated density of states.


\end{document}